\documentclass[english]{article}

\usepackage[colorlinks,bookmarksopen,bookmarksnumbered,allcolors=red]{hyperref}
\usepackage{xcolor}
\usepackage{graphicx}
\usepackage{geometry}
\usepackage{amsmath, amssymb, bm}
\usepackage{authblk}
\usepackage{placeins}
\usepackage{pdflscape}
\usepackage{subcaption}

\definecolor{bluegreen}{RGB}{46,141,131}
\definecolor{darkgreen}{RGB}{46,139,87}
\definecolor{darkred}{RGB}{219,7,61}
\definecolor{darkblue}{RGB}{0,0,137}

\title{\textbf{A Two-stage Joint Modeling Approach for Multiple Longitudinal Markers and Time-to-event Data}}

\author[,1]{Taban Baghfalaki\thanks{Corresponding author: taban.baghfalaki@u-bordeaux.fr}}
\author[2]{Reza Hashemi}
\author[1]{Catherine Helmer}
\author[1]{Helene Jacqmin-Gadda}
\affil[1]{Univ. Bordeaux, INSERM, INRIA, BPH, U1219, F-33000 Bordeaux, France}
\affil[2]{Department of Statistics, Razi University, Kermanshah, Iran}

\date{}

\begin{document}

\maketitle

\begin{abstract}
Collecting multiple longitudinal measurements and time-to-event outcomes is common in many clinical and epidemiological studies. Studying the associations between them is often the main focus of interest. There are several R packages available for implementing joint modeling of multiple longitudinal measurements and time-to-event outcomes, which is the most appropriate tool for analyzing this kind of data. However, when the number of longitudinal markers increases, estimating the joint model becomes increasingly difficult or even impossible due to very long computation times and convergence issues. In this paper, we propose a novel two-stage approach  in a Bayesian framework for estimating joint models for multiple longitudinal measurements and time-to-event outcomes. The proposed approach is similar to the standard two-stage approach, but at the first stage,  we estimate the  one-marker joint model for the event and each longitudinal marker instead of using mixed models. Using these estimates, predictions of expected values or slopes of the individual markers trajectories are obtained which avoid the bias due to informative dropouts. Then, at the second stage, a proportional hazard model is estimated including the expected current values and/or current slopes of all the markers as time-dependent covariates. To account for the uncertainty of the prediction at stage 1, a multiple imputation approach is used when estimating the Cox model at stage 2. This approach enables us to estimate prediction models based on numerous longitudinal markers which could be intractable with multi-marker joint modeling. The performance of the proposed approach for both parameter estimation and risk prediction is investigated through simulation studies and the use of the public PBC2 data set. Additionally, the proposed approach is applied to predict the risk of dementia using a real data set with seventeen longitudinal markers. To streamline implementation, we have developed an R package named \texttt{TSJM}, that is freely available on GitHub: \url{https://github.com/tbaghfalaki/TSJM}.
\end{abstract}

\section{Introduction}
In clinical trials and cohort studies, it is common to collect multiple longitudinal measurements along with the time to an event, such as death, recovery, or disease recurrence. For example, in the context of HIV infection, HIV-RNA and CD4 are often collected as dynamic biomarkers until death or censoring occurs \cite{guedj2011joint}. In most of the cohort studies and administrative health databases, the number of longitudinally collected markers may be very large. In modern medicine, there is often a need to estimate the association between the longitudinal trajectories of multiple biomarkers and the relative risks of a specific event \cite{hossain2020multivariate,shen2021backward,lin2022deep,devaux2022random,zhang2023multivariate}. This is particularly important for developing dynamic prediction models for the event using repeated measurements of the markers.
These longitudinal biomarkers are often collected with measurement errors and only at intermittent time points \cite{tsiatis1995modeling,dafni1998evaluating}.  As a result, these prediction models cannot be estimated with standard survival models since they require the true value of the markers at any time point (or at any time of event for the Cox model) \cite{ye2008semiparametric}.
Joint modeling of longitudinal and time-to-event data is an attractive approach for analyzing this type of dataset \cite{rizopoulos2012joint,elashoff2016joint}.
Contrary to the Cox model with time-dependent explanatory variables, joint models handle the measurement error and the discrete-time measurement process of the longitudinal marker. There is a significant amount of  literature on joint modeling and its extensions. For a review on joint modeling, see \cite{sousa2011review,papageorgiou2019overview,alsefri2020bayesian} and \cite{zhudenkov2022workflow}.
From a computational point of view, 
\texttt{JoineRML} \cite{hickey2018joinerml},
\texttt{JMBayes} \cite{rizopoulos2014r}, \texttt{JMBayes2} \cite{rizopoulos2022jmbayes2}, and \texttt{INLAjoint} \cite{rustand2022denisrustand} are now available in the R software. The implementation of joint modeling with a limited number of multiple markers is offered through \texttt{INLAjoint} \cite{rustand2022denisrustand} and  \texttt{JMBayes2} \cite{rizopoulos2022jmbayes2}. However, their computational difficulty becomes intractable when dealing with a large number of markers, e.g. more than ten.\\
As the estimation of joint models is challenging and time-consuming, two-stage approaches (also known as regression calibration or RC) have been proposed.
A standard two-stage (STS) approach is a conventional strategy for accounting  measurement errors in order to minimize the bias of the estimates \cite{self1992modeling,dafni1998evaluating,ye2008semiparametric,albert2010estimating}.
In the first stage of STS, a mixed effects model is fitted for the longitudinal marker without considering the time-to-event data. In the second stage, some functions of the posterior expectation of the individual's random effects from the mixed model are estimated and incorporated as time-dependent covariates in a Cox model. The STS approach and joint likelihood method for modeling longitudinal and survival data were compared via extensive simulations \cite{dafni1998evaluating,tsiatis2001semiparametric,sweeting2011joint,ye2017two,sayers2017joint}. They found that joint modeling outperforms two-step approaches, but the latter can be considered when there are large numbers of longitudinal markers and random effects. Efforts have been made to mitigate the bias resulting from the STS approach. Ye et al.\cite{ye2008semiparametric} proposed a two-stage RC approach in which the longitudinal data are modeled using a semiparametric mixed model. This approach allows for complex patterns in the longitudinal marker and is simpler to implement compared to a joint modeling approach. Although this approach mitigates the bias, it does not eliminate the bias  due to measurement error and informative dropout in the first stage estimation. To address the  issue of informative dropouts, Albert and  Shih \cite{albert2010estimating} proposed a modified RC approach. In this approach, the longitudinal outcomes are initially imputed using a pattern mixture model (PMM), and then the mixed model is estimated based on the imputed data set. This approach is indeed less biased than the standard RC approach. However, it becomes challenging to apply in practical situations when the time-to-event data is collected in continuous time and is censored. Additionally, if the time trend in the mixed model is not linear, it is necessary to have multiple measurements for each subject in order to estimate the PMM.
Huong et al. \cite{huong2018modified} proposed  maximizing the approximation of the full joint log-likelihood at the second stage.
An extension of their approach to cases with competing risks was discussed by Mehdizadeh et al.\cite{mehdizadeh2021two}. For the setting with multiple longitudinal biomarkers, Guler et al.\cite{guler2017two} extended the STS approach by considering pairwise modeling \cite{fieuws2004joint} in the first stage. Mauff et al.\cite{mauff2020joint} proposed a corrected STS for joint modeling of multiple longitudinal and time-to-event outcomes based on importance sampling.
Dutta and Chakraborty \cite{dutta2021joint} proposed a two-stage approach using a non-linear mixed-effects model for the first stage and  used   an adaptive gradient descent algorithm for the second stage.
However, none of the proposed two-stage approaches completely eliminates the bias caused by the informative dropout following the event. 
Additionally, Moreno-Betancur et al.\cite{moreno2018survival} introduced a method that addresses the complexities of integrating time-varying covariates into time-to-event models. This method utilizes multiple imputation for joint modeling, enabling the incorporation of multiple covariates into the time-to-event model.\\
In this paper, we propose a new two-stage approach for modeling multiple longitudinal markers and time-to-event data,  when the number of longitudinal markers is large and the estimation of the full joint model including all the markers is intractable. Using a Bayesian approach, in the first stage, the one-marker joint models are estimated for the event and each longitudinal marker  to account for the bias caused by informative dropout \cite{albert2010estimating}. In the second stage, a proportional hazard model with  time-dependent covariates is estimated for the time-to-event outcome, including 
as explanatory variables, the estimations of functions of the random effects and parameters from the first stage joint modeling of each longitudinal marker. \\
The main objective of the joint modeling of time-to-events and multiple longitudinal markers is the individual risk prediction. We  also propose formulas for computing these predictions using the two-stage approach. With a focus on enhancing user experience, we have developed the \texttt{TSJM} R package. Crafted to simplify implementation, it is readily accessible on GitHub at: \url{https://github.com/tbaghfalaki/TSJM}.
\\
This paper is organized as follows. In Section 2, we  first review the joint modeling for multiple longitudinal markers and time-to-event data. Then, we describe the STS approach and our proposed two-stage approach for modeling multiple longitudinal markers and a time-to-event outcome.  Section 3 focuses on dynamic prediction using the proposed approach. Section 4 presents a simulation study comparing the behavior of the proposed two-stage approach with the STS approach, the 2-stage approach proposed by Moreno-Betancur et al.\cite{moreno2018survival} and the multi-marker joint model (MMJM) in terms of parameter estimation and dynamic prediction.
In Section 5, the performance of the proposed approach is first compared with MMJM using the public PBC2 data. Then, it is applied to predict the risk of dementia in the Three-City (3C) cohort study \cite{antoniak2003vascular} using 17 longitudinal predictors. Finally, the last section of the paper includes a discussion of the advantages and limitations of the proposed approach.
\section{Multi-marker longitudinal and time-to-event outcome modeling }
In this section, we describe different methodologies, including MMJM, the STS approach and its extensions to the multi-marker setting, and our proposed two-stage joint modeling approach in a Bayesian paradigm.
\subsection{Multi-marker joint model (MMJM)}\label{sectionjm}

\subsubsection{Longitudinal sub-model}
Let $Y_{ijk}={Y}_{ik} (s_{ijk})$ be the $k$th longitudinal marker for the $i$th subject at time $s_{ijk}$. A multivariate linear mixed effects model is postulated for the multiple longitudinal markers as follows:
\begin{eqnarray}\label{m1}
  Y_{ijk} =  \eta_{ik}(s_{ijk}|\bm{\beta}_k,\bm{b}_{ik})+\varepsilon_{ijk}, i=1,~\ldots,N,~j=1,\ldots,n_i,~k=1,\ldots,K,
\end{eqnarray}
where $\varepsilon_{ijk}\overset{\text{iid}}{\sim} N(0, \sigma_k^2)$. We assume that 
\begin{eqnarray}\label{m2}
  \eta_{ik}(s|\bm{\beta}_k,\bm{b}_{ik}) = \bm{X}_{ik}(s)^\top \bm{\beta}_k + \bm{Z}_{ik}(s)^\top \bm{b}_{ik},~k=1,\ldots,K,
\end{eqnarray}
where $\bm{X}_{ik}(s)$ and $\bm{Z}_{ik}(s)$ are, respectively, the design vectors at time $s$ for the fixed effects $\bm{\beta}_k$ and the random effects $\bm{b}_{ik}$.\\
The association within and between the multiple longitudinal markers is taken into account by the random effects $\bm{b}_i=(\bm{b}_{i1},\ldots,\bm{b}_{iK})^\top$ such that
$\bm{b}_i \sim N(\bm{0},\bm{\Sigma})$, and $\bm{\Sigma}$ is $\sum_{k=1}^K p_k\times \sum_{k=1}^K p_k$ covariance matrix where $p_k$ is the size of $\bm{b}_{ik}$. Additionally, it is assumed that ${\varepsilon}_{ijk}$ is independent of $\bm{b}_{ik},~i=1,\ldots,K$.

\subsubsection{Time-to-event sub-model}
Let $T_{i}^*$ be the true failure time for subject $i$, such that the observed time is $T_i=min(T_{i}^*,C_i)$, where
$C_i$ denotes the censoring time. Also, $\delta_i$
is the failure indicator $\delta_i=1_{T_i^*<C_i}$.
The hazard of an event is assumed to be a function of the subject-specific linear predictor $\eta_{ik}(t|\bm{\beta}_k,\bm{b}_{ik})$ and/or the random effects in the time-to-event process as follows \cite{rizopoulos2012joint}:
\begin{eqnarray}\label{h}
\lambda_{i}(t|\bm{\omega}_i,\bm{b}_i)&=&\lambda_{0}(t)\exp\{\bm{\gamma}^\top \bm{\omega}_i +\sum_{k=1}^{K} 
g_k\big(\bm{b}_{ik},\bm{X}_{ik},\bm{Z}_{ik}, \bm{\alpha}_{k},\bm{t}\big) \},
\end{eqnarray}
where $\lambda_{0}(t)$
 denotes the baseline hazard function, $\bm{\omega}_i$
 is a vector of time-invariant covariates with corresponding regression coefficients $\bm{\gamma}$ and
 $g_k(.,.,.,.)$ is a known function 
 parameterized with the association parameter
 $\bm{\alpha}=(\bm{\alpha}_{1},\ldots,\bm{\alpha}_{K})^\top$ that
 specifies which components of the longitudinal marker process are included in the linear predictor of the hazard function. As some examples, we can use the current values $\big(\eta_{ik}(t|\bm{\beta}_k,\bm{b}_{ik}),~k=1,\ldots,K\big)$, the current slopes $\big(\frac{d {\eta_{ik}(t|\bm{\beta}_k,\bm{b}_{ik})}}{dt},~k=1,\ldots,K\big)$ or the random effects $\big(\bm{b}_{ik},~k=1,\ldots,K\big)$. For more details, refer to
\cite{rizopoulos2011bayesian,rizopoulos2012joint,taylor2013real,rizopoulos2014combining,rizopoulos2016personalized}.\\
 There are different approaches for specifying the baseline hazard function in the joint modeling \cite{rizopoulos2012joint}. Although, the methodology is not restricted to a specific form of the baseline hazard, here we consider
 a piecewise constant baseline hazard \cite{ibrahim2001bayesian}. Therefore,   a finite partition of the time axis  is constructed as $0=\xi_0<\xi_1<\xi_2<\ldots,\xi_J$, such that $max(T_1,\ldots,T_N)<\xi_J$. Thus, there are $J$ intervals as $(0,\xi_1],(\xi_1,\xi_2],\ldots,(\xi_{J-1},\xi_J]$. We consider a constant baseline hazard for each interval as $\lambda_0(t)=\lambda_j,~t\in (\xi_{j-1},\xi_j],~j=1,\ldots,J$. \\
Bayesian parameter estimation for this model is briefly presented  in the web Appendix A. 
\subsection{Standard two-stage approach}\label{sts}
The two-stage (TS) approach \cite{tsiatis1995modeling} is a useful strategy to approximate estimation of joint models and to reduce their computational time.
In the first stage, we estimate the mixed-effects model defined by \eqref{m1} and \eqref{m2} for each longitudinal marker, where $\bm{\Sigma}$ is considered to be block diagonal. We then estimate $\eta_{ik}(s|{\bm{\beta}}_k,\bm{b}_{ik})$ by using the posterior expectation of the individual's random effects from the mixed-effects models denoted by $\hat{\bm{b}}_{ik}$. In practice, $\hat{\bm{b}}_{ik}$  is computed as the empirical mean of the posterior distribution $(\bm{b}_{ik}|\bm{y}_{ik};{\bm{\beta}}_k,\sigma_k,\bm{\Sigma}_k)$,
where $\bm{\Sigma}_k$ represents the covariance matrix of the random effects  for the $k$th longitudinal marker.
Assume that $g_k\big(\bm{b}_{ik},\bm{X}_{ik},\bm{Z}_{ik}, \bm{\alpha}_{k}\big)=g_k\big({\eta}_{ik}(t|{\bm{\beta}}_k,{\bm{b}}_{ik}), \bm{\alpha}_{k}\big) $ in Equation \eqref{h}, that is, the hazard function is associated with a function of the current value.
Let
$\eta_{ik}(s|\hat{\bm{\beta}}_k,\hat{\bm{b}}_{ik})$ be the estimated linear predictor for the $k$th longitudinal model based on Equations \eqref{m1} and \eqref{m2}.
In the second stage, the values of $\eta_{ik}(t|\hat{\bm{\beta}}_k,\hat{\bm{b}}_{ik}),~k=1,\ldots,K$ are included as time-varying covariates in  a 
survival model. For this purpose, a proportional hazard model is considered as follows:
\begin{eqnarray}\label{hazz1}
\lambda_{i}(t|\bm{\omega}_i,\hat{\bm{b}}_i)=\lambda_{0}(t)\exp\{\bm{\gamma}^\top \bm{\omega}_i+\sum_{k=1}^{K} g_k\big(\eta_{ik}(t|\hat{\bm{\beta}}_k,\hat{\bm{b}}_{ik}), \bm{\alpha}_{k}\big) \}.
\end{eqnarray}
Most often, the PHM  used in phase 2 of the two-stage approach is semi- parametric (with undefined baseline risk function) for more flexibility. In the following, if $K=1$, we refer to it as standard TS (STS), and if $K>1$, we use the term multiple TS (MTS) to abbreviate this approach.

\subsection{Two-stage joint modeling accounting for informative dropout}\label{csts}
The MTS regression calibration approach may cause bias because estimated parameters from the mixed models at stage 1 are biased due to the fact that the event leads to informative missing longitudinal markers \cite{albert2010estimating}. To remove this bias, we propose replacing the estimation of the mixed models at the first stage with the estimation of $K$ one-marker joint models for the event and each marker. 
Therefore, our method comprises two distinct stages, collectively referred to as Two-Stage Joint Modeling (TSJM). The approach is outlined as follows:
\begin{description}
    \item[Stage 1] For each $k$, $k=1,\ldots,K$, estimate the joint model for the longitudinal marker $k$ and the time-to-event. The one-marker JMs are estimated using the joint posterior distribution A.2 for $K=1$. 
    Then, predict the random effects $\hat{\bm{b}}_{ik}$ and the linear predictor $\eta_{ik}(t|\hat{\bm{\beta}}_k,\hat{\bm{b}}_{ik})$, for marker $k$ using the empirical mean of the posterior distribution $(\bm{b}_{ik}|\bm{y}_{ik},t_i,\delta_i,\hat{\bm{\theta}}_k)$, where $\hat{\bm{\theta}}_k$ represents the estimated parameters for the one-marker joint model for the $k$th longitudinal marker. 
This first stage allow the estimation of $\eta_{ik}(t|\hat{\bm{\beta}}_k,\hat{\bm{b}}_{ik})$ without the bias due to informative dropout. 

  \item[Stage 2] In the second stage, similar to the MTS approach, the values of $\eta_{ik}(t|\hat{\bm{\beta}}_k,\hat{\bm{b}}_{ik})$ are included as time-varying covariates in a Cox model to estimate parameters $\bm{\gamma}$ and $
  \bm{\alpha}_k$, as well as the baseline hazard from \eqref{hazz1} (we call them $\Tilde{\bm{\gamma}}$ and $
  \Tilde{\bm{\alpha}}_k$ $k=1,\cdots,K$). 
To estimate the parameters in the Cox model with time-dependent covariates \eqref{hazz1}, we maximize the log-partial likelihood \cite{therneau2017using,lin2002modeling}. The log-partial likelihood for the model is
\begin{eqnarray*}
\ell_{{\it{P}}}(\bm{\gamma}, \bm{\alpha}_k,~k=1,\cdots,K) &=& \sum_{i:\delta_i=1} \left\{ \bm{\gamma}^\top \bm{\omega}_i + \sum_{k=1}^{K} g_k(\eta_{ik}(t_i|\hat{\bm{\beta}}_k, \hat{\bm{b}}_{ik}), \bm{\alpha}_k) \right. \nonumber \\
&& \left. - \log \left( \sum_{j \in R(t_i)} \exp(\bm{\gamma}^\top \bm{\omega}_j + \sum_{k=1}^{K} g_k(\eta_{jk}(t_i|\hat{\bm{\beta}}_k, \hat{\bm{b}}_{jk}), \bm{\alpha}_k)) \right) \right\}.
\end{eqnarray*}
Here, \( R(t_i) \) is the risk set at time \( t_i \), which includes all individuals still at risk just before time \( t_i \). The parameter estimates \(\bm{\gamma}\) and \(\bm{\alpha}_k,~k=1,\cdots,K\) 
are obtained  iteratively  using Newton-Raphson optimization. This process continues until convergence is achieved and can be performed using the \texttt{survival} package \cite{therneau2013r} in R, which supports the inclusion of time-varying covariates.\\
To account for the uncertainty of the estimates of stage 1, the standard errors of the parameters from the Cox model \eqref{hazz1} are estimated using the Rubin's formula for multiple imputation  \cite{little1987multiple}. We consider \( \mathcal{M} \) realizations of \( \bm{\theta}_k \) and \( \bm{b}_{ik} \) sampled from their posterior distribution at stage 1, and we re-estimate the parameters from the Cox model for each of these \( \mathcal{M} \) realizations. Denoting $\Tilde{\vartheta}_m=(\bm{\Tilde{\gamma}}_m, \bm{\Tilde{\alpha}}_{km})$ the estimates of parameters from model \eqref{hazz1} for the imputed data set $m$, the combined estimate is obtained by 
\[
\bar{\vartheta}_{\mathcal{M}} = \frac{1}{\mathcal{M}} \sum_{m=1}^{\mathcal{M}} \Tilde{\vartheta}_m.
\] 
The average within-imputation variance \( \bar{W}_{\mathcal{M}} \) is 
\[
\bar{W}_{\mathcal{M}} = \frac{1}{\mathcal{M}} \sum_{m=1}^{\mathcal{M}} W_m,
\]
where in our approach, \( W_m \) is computed based on the inverse of the Hessian matrix. The between-imputation variance \( B_{\mathcal{M}} \) is 
\[
B_{\mathcal{M}} = \frac{1}{\mathcal{M}-1} \sum_{m=1}^{\mathcal{M}} (\Tilde{\vartheta}_m - \bar{\vartheta}_{\mathcal{M}})^2.
\] 
The total variability \( T_{\mathcal{M}} \) associated with the combined estimate is 
\[
T_{\mathcal{M}} = \bar{W}_{\mathcal{M}} + \left(1 + \frac{1}{\mathcal{M}}\right) B_{\mathcal{M}}.
\] 
 \end{description}
\section{Dynamic predictions from TSJM}
\subsection{Individual risk predictions}\label{indrisk}
In the joint modeling of longitudinal measurements and time-to-event outcomes, there has been significant attention given to the development of dynamic prediction (DP). DP refers to the risk predictions for subject $i$ who has survived up to time point $s$ and has provided longitudinal measurements $\bm{\mathcal{Y}}_{i}(s)$.
Let the proposed joint models be fitted to a random sample $\bm{\mathcal{D}}=\{\bm{Y}_{i1},\ldots,\bm{Y}_{iK},T_i,\delta_i,~i=1,\ldots,N \}$ which is called training data. In clinical practice and for validation purpose, we are interested in predicting the survival probability of subjects not included in the learning set. For each subject $i$ in this new set (thereafter denoted validation set), we have access to longitudinal measurements $\bm{\mathcal{Y}}_{i}(s)$ given by $\{Y_{ik}(s_{ijk}); 0\le s_{ijk}\le s, j=1,\cdots,n_i, k=1,\cdots,K \}$ and the vector of available explanatory variables $\bm{\mathcal{X}}_{i}(s)$ given by $\{\bm{\omega}_i, \bm{X}_{ijk}; 0\le s_{ijk}\le s, j=1,\cdots,n_i, k=1,\cdots,K \}$.
Then, for the prediction time $s$ and the prediction window $t$, the risk probability is given by:
\begin{eqnarray}\label{dp}
\pi_i(s+t \mid s) & = &\mathrm{P}\left(s \leq T_i^*<s+t \mid T_i^*>s, \mathcal{Y}_i(s), \boldsymbol{\mathcal { X }}_i(s) ; \boldsymbol{\vartheta},\boldsymbol{\theta}\right) \\\nonumber
& =& 1-\mathrm{P}\left(T_i^*>s+t \mid T_i^*>s, \mathcal{Y}_i(s), \boldsymbol{\mathcal { X }}_i(s) ; \boldsymbol{\vartheta},\boldsymbol{\theta}\right)\nonumber
\end{eqnarray}
where $\bm{\vartheta}$ is the  unknown parameters of the Cox model and $\boldsymbol{\theta}=(\boldsymbol{\theta}_1,
\cdots,\boldsymbol{\theta}_K)$ is the unknown parameters of the longitudinal markers at the first stage, and $T_i^*$ denotes the true event time for the $i$th subject.
There are various strategies for estimating this quantity. Note that we have
\begin{eqnarray}\label{dppp}
\mathrm{P}\left(T_i^*>s+t \mid T_i^*>s, \mathcal{Y}_i(s), \boldsymbol{\mathcal { X }}_i(s) ; \boldsymbol{\vartheta}\right)  & = &\int \frac{S\left(s+t \mid \boldsymbol{b}_i, \boldsymbol{\mathcal { X }}_i(s) ; \boldsymbol{\vartheta},\boldsymbol{\theta}\right)}{S\left(s \mid \boldsymbol{b}_i, \boldsymbol{\mathcal { X }}_i(s) ; \boldsymbol{\vartheta},\boldsymbol{\theta}\right)} \\\nonumber
& \times & f\left(\boldsymbol{b}_i \mid T_i^*>s, \mathcal{Y}_i(s), \boldsymbol{\mathcal { X }}_i(s) ; \boldsymbol{\vartheta},\boldsymbol{\theta}\right) d \boldsymbol{b}_i,\nonumber
\end{eqnarray}
where $S(\cdot)$ represents the survival function conditional on the random effects.
For estimating this quantity, first of all, the parameters $\bm{\vartheta}$ must be estimated based on the training data $\bm{\mathcal{D}}$ in the second stage. This estimates is denoted $\Tilde{\bm{\vartheta}}$.
A Monte Carlo scheme and a first-order approximation of it are proposed by \cite{rizopoulos2011dynamic} to obtain estimates of these probabilities.
 We adapt both approaches to the TSJM estimation procedure. The first-order approximation of ${\mathbb{\pi}} _{i}(s+t|s)$ is as follows:
\begin{eqnarray}\label{ap2}
1-\hat{\mathbb{\pi}} _{i}(s+t|s)  &= &Pr(T^*_i > s+t | T^*_i > s, \hat{\bm{b}}_i, \bm{\mathcal{X}}_{i}(s); \Tilde{\bm{\vartheta}},\hat{\bm{\theta}})\\ &= &
 \frac{S(s+t|\hat{\bm{b}}_i,\bm{\mathcal{X}}_{i}(s);\Tilde{\bm{\vartheta}},\hat{\bm{\theta}})}{S(s| \hat{\bm{b}}_i,\bm{\mathcal{X}}_{i}(s);\Tilde{\bm{\vartheta}},\hat{\bm{\theta}})},\nonumber
\end{eqnarray}
where $\hat{\bm{b}}_i$ represents the predicted random effects, which are based on the empirical mean of the distribution $p(\bm{b}_{i}|T^*_i > s,\bm{\mathcal{Y}}_{i}(s),\bm{\mathcal{X}}_{i}(s);\hat{\bm{\theta}})$. For the TSJM, it is not possible to draw a sample directly from this distribution. Instead, we consider the estimated parameters $\hat{\bm{\theta}}_k$ obtained from the training data and the estimation of the one-marker JM based on the $k$th marker. We then generate $L$ random samples from the distribution of $\bm{b}_{ik}$ given $(T^*_i > s,\bm{\mathcal{Y}}_{i}(s),\bm{\mathcal{X}}_{i}(s), \hat{\bm{\theta}}_k)$, and refer to them as $\bm{b}_{ik}^{(\ell)},~\ell=1,\cdots,L$. We, then, merge the predicted random effects to obtain the $\sum_{k=1}^Kp_k$-dimensional predicted random effect $\bm{b}_{i}^{(\ell)}=(\bm{b}_{i1}^{(\ell)},\ldots,\bm{b}_{iK}^{(\ell)}),~\ell=1,\cdots,L$. Finally, we calculate the average of the generated samples to obtain $\hat{\bm{b}}_{i}$. As a summary, for the TSJM, we can follow the algorithm outlined below:\\
{\bf{Algorithm 1:}} First-order approximation of DP.
    \begin{enumerate}
    \item Consider the one-marker joint models for each marker and predict the $p_k$-dimensional random effect
$\bm{b}_{ik}$ given  $(T^*_i > s,\bm{\mathcal{Y}}_{i}(s),\bm{\mathcal{X}}_{i}(s), \hat{\bm{\theta}}_k)$, denoted by $\hat{\bm{b}}_{ik}$. Note that $\hat{\bm{\theta}}_k$ represents the estimated parameters for the one-marker JM  in the first stage of TSJM.
\item Merge the predicted random effects from the previous step to obtain the $\sum_{k=1}^Kp_k$-dimensional predicted random effect $\hat{\bm{b}}_{i}=(\hat{\bm{b}}_{i1},\ldots,\hat{\bm{b}}_{iK})$.
\item Compute \eqref{ap2} using the vector $\hat{\bm{b}}_{i}$ and the estimated parameter $\Tilde{\bm{\vartheta}}$ using TSJM.\\
    \end{enumerate}
In practice, deriving the standard errors and confidence intervals for the first-order approximation of DP is a rather difficult task \cite{rizopoulos2011dynamic}. To compute the standard error of the DP, another algorithm, similar to the proposed methods of Rizopoulos \cite{rizopoulos2011dynamic} and Proust-Lima and Taylor \cite{proust2009development} can be considered as follows:\\
\\
{\bf{Algorithm 2:}} Monte-Carlo approximation of DP.
    \begin{enumerate}
    \item Consider a realization of the generated MCMC sample for estimating the unknown parameters from the one-marker JM for marker $k$ in the training data, denoted $\bm{\theta}_k^{(\ell)},~\ell=1,\ldots,L$.
    \item Draw $M$ values $\bm{b}_{ik}^{(\ell,m)}$ from the posterior distribution $(\bm{b}_{ik} | T^*_i > s,\bm{\mathcal{Y}}_{i}(s),\bm{\mathcal{X}}_{i}(s), \bm{\theta}_k^{(\ell)})$. 
    \item  Let $\overline{\bm{b}}_i^{(\ell)}=\frac{1}{M}\sum_{m=1}^{M} \bm{b}_i^{(\ell,m)}$, then compute the following quantity: 
    \begin{eqnarray*}
1-\hat{\mathbb{\pi}} _{i}(s+t|s)^{(\ell)} =\frac{S(s+t|\overline{\bm{b}}_i^{(\ell)},\bm{\mathcal{X}}_{i}(s);\Tilde{\bm{\vartheta}},\hat{\bm{\theta}})}{S(s| \overline{\bm{b}}_i^{(\ell)},\bm{\mathcal{X}}_{i}(s);\Tilde{\bm{\vartheta}},\hat{\bm{\theta}})},~\ell=1,\ldots,L.
\end{eqnarray*}
    \item  Compute summary statistics, including mean, standard error, percentiles, etc., based on $L$ generated samples of Step 3.
\end{enumerate}
In practice, Algorithm 1 is efficient and recommended for use. However, the second Algorithm is more suitable for obtaining standard errors and credible intervals.
\subsection{Measures of prediction accuracy}
The Area Under the Receiver Operating Characteristics Curve (AUC) and the Brier Score (BS) are two criteria used to evaluate the predictive abilities of prediction models \cite{steyerberg2009applications}.
The AUC measures the discrimination between future diseased and non-diseased individuals. The AUC represents the probability that the predicted risk of a randomly selected diseased subject is greater than the predicted risk of a disease-free subject. Also, BS quantifies the accuracy of the predictions, specifically the mean squared error of the predictions.
We use the non-parametric inverse probability of censoring weighted (IPCW)
estimators of AUC and BS for dealing with  censored time-to-event and time-dependent markers  \cite{blanche2015quantifying}.

\section{Simulation study}
A simulation study was conducted to assess the effectiveness of the proposed TSJM approach and to compare it with the multi-marker joint modeling (MMJM) approach, MTS and the 
 two stage approach proposed by Moreno-Betancur et al.\cite{moreno2018survival}  which is available in R package \texttt{survtd} \cite{moreno2017package}  (denoted MB) that also aims at avoiding bias due to informative dropouts and measurement error.
The simulation scenarios varied according to the strength of the association between the markers and the event which influences the intensity of the informative dropout bias in MTS, the residual variances for the markers that quantify measurement error and may impact performances of both TS approaches, and the correlation between the markers that is neglected in the first stage of the TS approaches but partly handled in MB. In addition to the criteria for parameter estimation, such as root mean square error (RMSE), relative bias (RB), and coverage rate of the $95\%$ credible interval (CR), the performances of the approaches are also evaluated through their predictive abilities using AUC and BS.
\subsection{Design of simulation}
Data are generated using multi-marker joint models with $K=4$ longitudinal markers and a survival model. To be able to compare the performance to that of MMJM and limit  computation time, the simulation scenario considered only four markers.
More specifically, the data generation models were as follows:
\begin{eqnarray}\label{gau}
Y_{ik}(t)&=&\eta_{ik}(t|\bm{\beta}_k,\bm{b}_{ik})+\varepsilon_{ikt}\\
&=&\beta_{0k}+\beta_{1k}t+\beta_{2k}x_{1i}+\beta_{3k}x_{2i}+b_{0ki}+b_{1ki} t+\varepsilon_{ikt},\nonumber
\end{eqnarray}
where $\varepsilon_{ikt} \sim N(0, \sigma_k^2)$,
$x_1$ is generated from a binary distribution with a success probability of 0.5, and $x_2$ is generated from a normal distribution with a mean of zero and a standard deviation of 0.5. Also, $\bm{b}_i=(b_{01i},b_{11i},\ldots,b_{04i},b_{14i})^{\top} \sim N(\bm{0},\bm{\Sigma})$, $i=1,\ldots,n$ and $t=0,0.2,0.4,\ldots,2$.\\
The time-to-event outcome was simulated using a proportional hazards model with a permutational algorithm \cite{abrahamowicz1996time,sylvestre2008comparison}, for generating data with time-dependent covariates. In our case, it depends on the current values of the $K$ markers as follows:
$$\lambda(t)=\lambda_{0}(t) \exp (\sum_{k=1}^K \alpha_{k} \eta_{ik}(t|\bm{\beta}_k,\bm{b}_{ik})),$$
This mechanism led to approximately 70\% of the sample being censored and a median number of measurements of 5  per subjects.\\
For the covariance matrix of the random effects a structure is considered as follows:
\begin{eqnarray}\label{c}
\bm{\Sigma}=\begin{pmatrix}
\bm{\Sigma}^* & \bm{\Sigma}^\dagger & \bm{\Sigma}^\dagger & \bm{\Sigma}^\dagger \\
  & \bm{\Sigma}^* & \bm{\Sigma}^\dagger & \bm{\Sigma}^\dagger \\
  &  & \bm{\Sigma}^* & \bm{\Sigma}^\dagger \\
 &   &   & \bm{\Sigma}^* 
\end{pmatrix}, ~~ \rm{where}~~
\bm{\Sigma}^*=\begin{pmatrix}
1 & \rho^*\\
\rho^*  & 1  
\end{pmatrix}, ~\bm{\Sigma}^\dagger=\rho^\dagger\begin{pmatrix}
1 & 1\\
1 & 1  
\end{pmatrix}.
\end{eqnarray}
The real values of the parameters are $\bm{\beta}_k=(-0.5,0.5,0.5,0.5)^\prime,~k=1,\ldots,4$.
To assess the influence of increasing measurement errors, association parameters, and marker correlation on the behavior of the three methods, we examined various values for these parameters, as outlined below:
\begin{enumerate}
   \item $\sigma_k^2=0.5,~1,~2$ for $k=1,\ldots,4$, 
   \item $\bm{\alpha}=(-0.2,-0.2,0.2,0.2)^\top,~(-0.5,-0.5,0.5,0.5)^\top$,
   \item $(\rho^\dagger,\rho^*)^\prime=(0.1,0.5)^\prime,~(0.4,0.5)^\prime,~(0.6,0.8)^\prime$.
\end{enumerate} 
For each scenario, $100$ data sets of $n=1000$ subjects were generated and split into a learning sample of $500$ subjects ($50\%$) and a validation sample of $500$ subjects.
The investigation focused on the ability to estimate parameters on the learning sample using  MMJM, TSJM, MTS and MB.
The performance of these approaches for risk prediction was then evaluated using the validation samples for MMJM, TSJM and MTS.\\
Using a MacBook Pro 2020 equipped with Apple's advanced M1 chip and 256GB of storage capacity, the mean (SD) computational times for MMJM, TSJM, MTS, and MB are 51.591 (9.379), 45.295 (4.809), 1.150 (0.066), and 1.371 (0.093), respectively. We implemented the TSJM without parallelization, and it is clear that parallelization effectively reduces computational time. \\ 
Tables \ref{sum.1}, \ref{sum.4} and \ref{sum.6} provides the RMSE, relative bias  and coverage rate of the credible interval for the association parameters with the four methods and the TRUE models and for $(\rho^\dagger,\rho^*)^\prime=(0.1,0.5)^\prime,~(0.4,0.5)^\prime$ and $(0.6,0.8)^\prime$, respectively. The TRUE model involves estimating association parameters by using both the real value of the parameters and the random effects from the mixed model \eqref{gau}.
On average, nearly 30\% of the generated samples  did not converge for the MB approach. Therefore, the results for this approach are based only on the samples that did converge.
The results indicate that the performance of TSJM is comparable to that of MMJM. While the MTS approach performs well for small values of $\alpha_k$ and $\sigma_k^2$ ($k=1,\ldots,K$), its performance declines when the association between the markers and the event risk or the measurement error increase. Comparing TSJM with the MB approach, we find that although the coverage rate for MB is higher, the bias and RMSE for MB are larger than those for TSJM particularly when the measurement error increases. Additionally, as the correlation between outcomes increases, the performance of all approaches decreases. Of note, in some cases, the RMSE of MMJM is higher than the one of TSJM.  \\
Tables \ref{auc.1}, \ref{auc.4}, and \ref{auc.6} present  the performance of the approaches for risk prediction. Note that since the purpose of the MB approach is parameter estimation and the available R package \texttt{survtd} \cite{moreno2017package} only performs parameter estimation, we  exclude the MB approach from this comparison. These tables include the mean (SD) of AUC and BS for a validation set across various values of $\bm{\alpha}$ and $\rho$. The results show that the predictive abilities of models estimated by TSJM are comparable to those of the TRUE model. 
In contrast, the performance of MTS is consistently poorer than that of TSJM and deteriorates further with increased measurement error or stronger associations between event risk and the markers.
Additionally, the values of the  BS for the MMJM are worse for large values of the association parameters, even though the results based on the AUC are comparable to those obtained by the TRUE and TSJM methods. One possible reason for this discrepancy is the use of a semi-parametric Cox model for the survival sub-model in the TS approaches, whereas the baseline risk is assumed to be piecewise constant in MMJM. Moreover, convergence issues may also impact the performances of MMJM.

\begin{landscape}
\begin{table}
 \centering
 \footnotesize
 \caption{\label{sum.1} Results of the simulation study including root mean square error (RMSE), relative bias (RB), and coverage rate (CR) for the estimates of the association parameters between event risk and the current value of the four longitudinal markers. The study is based on 100 simulated datasets, each with a sample size of 500, and $(\rho^\dagger,\rho^*)' = (0.1, 0.5)'$.}
\centering
\begin{tabular}{c|ccc|ccc|ccc|ccc|ccc}
  \hline
 &\multicolumn{3}{c|}{MMJM } &\multicolumn{3}{c|}{TSJM} &\multicolumn{3}{c|}{MTS}&\multicolumn{3}{c|}{MB}&\multicolumn{3}{c}{TRUE } \\    \hline
&  RMSE & RB & CR & RMSE & RB & CR & RMSE & RB & CR & RMSE & RB & CR & RMSE & RB & CR \\ 
  \hline
  & & \multicolumn{12}{c}{$\bm{\alpha}=(-0.2,-0.2,0.2,0.2)^\top$, $\sigma_k^2=0.5,~k=1,\ldots,4$}\\
      \hline
$\alpha_1$  & 0.070 & -0.021 & 0.960 & 0.078 & 0.066 & 0.940 & 0.068 & -0.049 & 0.950 & 0.082 & -0.061 & 0.987 & 0.063 & 0.033 & 0.980 \\ 
$\alpha_2$  & 0.062 & 0.028 & 0.940 & 0.071 & 0.121 & 0.940 & 0.059 & -0.003 & 0.970   & 0.067 & -0.045 & 0.974 & 0.065 & 0.054 & 0.960 \\ 
$\alpha_3$ & 0.063 & 0.007 & 0.970 & 0.071 & 0.087 & 0.970 & 0.061 & -0.033 & 0.980 & 0.064 & -0.006 & 0.987 & 0.061 & 0.039 & 0.980 \\ 
$\alpha_4$ & 0.073 & 0.013 & 0.910 & 0.081 & 0.096 & 0.910 & 0.070 & -0.024 & 0.960 & 0.086 & -0.071 & 0.934 & 0.073 & 0.064 & 0.910 \\ 
  & & \multicolumn{12}{c}{$\bm{\alpha}=(-0.2,-0.2,0.2,0.2)^\top$, $\sigma_k^2=1,~k=1,\ldots,4$}\\
$\alpha_1$ & 0.076 & -0.019 & 0.960 & 0.086 & 0.091 & 0.920 & 0.071 & -0.104 & 0.950 & 0.093 & -0.078 & 0.987 & 0.063 & 0.033 & 0.980 \\ 
$\alpha_2$ & 0.066 & 0.046 & 0.940 & 0.079 & 0.163 & 0.940 & 0.059 & -0.045 & 0.970 & 0.077 & -0.061 & 0.947 & 0.065 & 0.054 & 0.960 \\ 
$\alpha_3$  & 0.067 & 0.018 & 0.970 & 0.078 & 0.121 & 0.950 & 0.063 & -0.081 & 0.980 & 0.073 & -0.007 & 0.987 & 0.061 & 0.039 & 0.980 \\ 
$\alpha_4$  & 0.076 & 0.018 & 0.930 & 0.088 & 0.119 & 0.910 & 0.072 & -0.079 & 0.950 & 0.097 & -0.101 & 0.934 & 0.073 & 0.064 & 0.910 \\ 
  & & \multicolumn{12}{c}{$\bm{\alpha}=(-0.2,-0.2,0.2,0.2)^\top$, $\sigma_k^2=2,~k=1,\ldots,4$}\\
$\alpha_1$  & 0.086 & -0.023 & 0.920 & 0.091 & 0.130 & 0.910 & 0.078 & -0.179 & 0.920 & 0.107 & -0.094 & 1.000 & 0.063 & 0.033 & 0.980 \\ 
$\alpha_2$ & 0.075 & 0.068 & 0.960 & 0.096 & 0.235 & 0.950 & 0.062 & -0.102 & 0.980 & 0.096 & -0.083 & 0.932 & 0.065 & 0.054 & 0.960  \\ 
$\alpha_3$ & 0.075 & 0.026 & 0.960 & 0.095 & 0.176 & 0.920 & 0.069 & -0.147 & 0.980  & 0.090 & -0.003 & 1.000 & 0.061 & 0.039 & 0.980 \\ 
$\alpha_4$ & 0.083 & 0.022 & 0.930 & 0.101 & 0.164 & 0.890 & 0.076 & -0.153 & 0.930 & 0.114 & -0.147 & 0.932 & 0.073 & 0.064 & 0.910 \\ 
  & & \multicolumn{12}{c}{$\bm{\alpha}=(-0.5,-0.5,0.5,0.5)^\top$, $\sigma_k^2=0.5,~k=1,\ldots,4$}\\
$\alpha_1$  & 0.079 & 0.019 & 0.960 & 0.082 & 0.056 & 0.960 & 0.073 & -0.053 & 0.980  & 0.085 & -0.056 & 0.974 & 0.070 & 0.037 & 0.940 \\ 
$\alpha_2$ & 0.083 & 0.034 & 0.940 & 0.089 & 0.076 & 0.930 & 0.073 & -0.034 & 0.960  & 0.083 & -0.010 & 1.000 & 0.076 & 0.064 & 0.940 \\ 
$\alpha_3$ & 0.079 & 0.030 & 0.970 & 0.083 & 0.065 & 0.950 & 0.072 & -0.048 & 0.970 & 0.077 & -0.014 & 1.000 & 0.074 & 0.043 & 0.950 \\ 
$\alpha_4$ & 0.092 & 0.035 & 0.910 & 0.095 & 0.070 & 0.910 & 0.080 & -0.039 & 0.920  & 0.080 & -0.020 & 0.987 & 0.086 & 0.057 & 0.880 \\ 
  & & \multicolumn{12}{c}{$\bm{\alpha}=(-0.5,-0.5,0.5,0.5)^\top$, $\sigma_k^2=1,~k=1,\ldots,4$}\\
$\alpha_1$ & 0.093 & 0.042 & 0.940 & 0.089 & 0.059 & 0.930 & 0.090 & -0.119 & 0.900  & 0.101 & -0.090 & 0.974 & 0.070 & 0.037 & 0.940 \\ 
$\alpha_2$  & 0.095 & 0.052 & 0.920 & 0.096 & 0.075 & 0.900 & 0.088 & -0.104 & 0.900  & 0.096 & -0.043 & 1.000 & 0.076 & 0.064 & 0.940 \\ 
$\alpha_3$  & 0.091 & 0.051 & 0.960 & 0.087 & 0.064 & 0.960 & 0.089 & -0.119 & 0.890  & 0.089 & -0.039 & 1.000 & 0.074 & 0.043 & 0.950 \\ 
$\alpha_4$ & 0.107 & 0.060 & 0.890 & 0.102 & 0.071 & 0.900 & 0.093 & -0.107 & 0.870 & 0.088 & -0.053 & 0.987 & 0.086 & 0.057 & 0.880 \\
       & & \multicolumn{12}{c}{$\bm{\alpha}=(-0.5,-0.5,0.5,0.5)^\top$, $\sigma_k^2=2,~k=1,\ldots,4$}\\
$\alpha_1$  & 0.121 & 0.080 & 0.930 & 0.101 & 0.064 & 0.900 & 0.124 & -0.209 & 0.790 & 0.128 & -0.142 & 0.961 & 0.070 & 0.037 & 0.940 \\ 
$\alpha_2$ & 0.128 & 0.088 & 0.880 & 0.109 & 0.076 & 0.880 & 0.122 & -0.199 & 0.770 & 0.118 & -0.094 & 1.000 & 0.076 & 0.064 & 0.940 \\ 
$\alpha_3$ & 0.116 & 0.082 & 0.970 & 0.095 & 0.065 & 0.950 & 0.125 & -0.214 & 0.740  & 0.110 & -0.081 & 0.987 & 0.074 & 0.043 & 0.950 \\ 
$\alpha_4$ & 0.145 & 0.108 & 0.900 & 0.115 & 0.078 & 0.880 & 0.125 & -0.199 & 0.770 & 0.110 & -0.105 & 0.974 & 0.086 & 0.057 & 0.880 \\ 
 \hline
\end{tabular}
\end{table}
\end{landscape}

\begin{landscape}
\begin{table}
 \centering
 \footnotesize
 \caption{\label{auc.1} Results of the simulation study including AUC and BS for 100 simulated datasets, each with a sample size of 500 and $(\rho^\dagger,\rho^*)' = (0.1, 0.5)'$.}
\centering
\begin{tabular}{c|cccccccc|cccccccc}
  \hline
 &\multicolumn{8}{c|}{AUC} &\multicolumn{8}{c}{BS } \\    \hline
  &\multicolumn{2}{c}{MMJM } &\multicolumn{2}{c}{TSJM} &\multicolumn{2}{c}{MTS}&\multicolumn{2}{c|}{TRUE } & \multicolumn{2}{c}{MMJM } &\multicolumn{2}{c}{TSJM} &\multicolumn{2}{c}{MTS}&\multicolumn{2}{c}{TRUE } \\    \hline
& mean & sd & mean & sd & mean & sd & mean & sd & mean & sd & mean & sd & mean & sd & mean & sd  \\   \hline
  & & \multicolumn{12}{c}{$\bm{\alpha}=(-0.2,-0.2,0.2,0.2)^\top$, $\sigma_k^2=0.5,~k=1,\ldots,4$}\\
0 & 0.622 & 0.049 & 0.614 & 0.051 & 0.584 & 0.053 & 0.613 & 0.049 & 0.062 & 0.012 & 0.062 & 0.009 & 0.062 & 0.009 & 0.061 & 0.009 \\ 
0.25 & 0.622 & 0.060 & 0.620 & 0.060 & 0.601 & 0.061 & 0.615 & 0.054 & 0.068 & 0.010 & 0.068 & 0.010 & 0.069 & 0.010 & 0.068 & 0.010 \\ 
0.5 & 0.646 & 0.064 & 0.645 & 0.071 & 0.627 & 0.071 & 0.638 & 0.066 & 0.062 & 0.011 & 0.062 & 0.011 & 0.062 & 0.011 & 0.061 & 0.011 \\ 
0.75 & 0.667 & 0.071 & 0.664 & 0.072 & 0.646 & 0.075 & 0.656 & 0.069 & 0.066 & 0.014 & 0.070 & 0.016 & 0.070 & 0.015 & 0.069 & 0.014 \\ 
    & & \multicolumn{12}{c}{$\bm{\alpha}=(-0.2,-0.2,0.2,0.2)^\top$, $\sigma_k^2=1,~k=1,\ldots,4$}\\
0 & 0.632 & 0.051 & 0.619 & 0.051 & 0.569 & 0.055 & 0.613 & 0.049 & 0.065 & 0.021 & 0.062 & 0.009 & 0.062 & 0.009 & 0.061 & 0.009 \\ 
0.25 & 0.629 & 0.062 & 0.624 & 0.063 & 0.590 & 0.063 & 0.615 & 0.054 & 0.069 & 0.013 & 0.069 & 0.010 & 0.069 & 0.010 & 0.068 & 0.010 \\ 
0.5 & 0.651 & 0.066 & 0.649 & 0.072 & 0.619 & 0.071 & 0.638 & 0.066 & 0.062 & 0.012 & 0.062 & 0.012 & 0.062 & 0.012 & 0.061 & 0.011 \\ 
0.75 & 0.670 & 0.070 & 0.665 & 0.073 & 0.638 & 0.075 & 0.656 & 0.069 & 0.066 & 0.014 & 0.071 & 0.016 & 0.070 & 0.015 & 0.069 & 0.014 \\ 
   & & \multicolumn{12}{c}{$\bm{\alpha}=(-0.2,-0.2,0.2,0.2)^\top$, $\sigma_k^2=2,~k=1,\ldots,4$}\\
0 & 0.648 & 0.052 & 0.631 & 0.053 & 0.552 & 0.056 & 0.613 & 0.049 & 0.069 & 0.032 & 0.062 & 0.009 & 0.062 & 0.009 & 0.061 & 0.009 \\ 
0.25 & 0.643 & 0.066 & 0.631 & 0.065 & 0.575 & 0.065 & 0.615 & 0.054 & 0.071 & 0.023 & 0.069 & 0.010 & 0.069 & 0.010 & 0.068 & 0.010 \\ 
0.5 & 0.660 & 0.066 & 0.660 & 0.073 & 0.606 & 0.072 & 0.638 & 0.066 & 0.063 & 0.022 & 0.062 & 0.012 & 0.063 & 0.012 & 0.061 & 0.011 \\ 
0.75 & 0.675 & 0.072 & 0.667 & 0.074 & 0.626 & 0.076 & 0.656 & 0.069 & 0.067 & 0.016 & 0.072 & 0.017 & 0.070 & 0.015 & 0.069 & 0.014 \\ 
    & & \multicolumn{12}{c}{$\bm{\alpha}=(-0.5,-0.5,0.5,0.5)^\top$, $\sigma_k^2=0.5,~k=1,\ldots,4$}\\
0 & 0.640 & 0.098 & 0.745 & 0.050 & 0.709 & 0.051 & 0.759 & 0.040 & 0.219 & 0.095 & 0.066 & 0.022 & 0.061 & 0.010 & 0.058 & 0.009 \\ 
0.25 & 0.701 & 0.094 & 0.780 & 0.046 & 0.751 & 0.051 & 0.788 & 0.049 & 0.178 & 0.096 & 0.065 & 0.015 & 0.065 & 0.011 & 0.060 & 0.009 \\ 
0.5 & 0.752 & 0.086 & 0.807 & 0.056 & 0.779 & 0.057 & 0.805 & 0.062 & 0.133 & 0.084 & 0.057 & 0.011 & 0.059 & 0.011 & 0.053 & 0.010 \\ 
0.75 & 0.793 & 0.085 & 0.814 & 0.063 & 0.786 & 0.071 & 0.810 & 0.066 & 0.105 & 0.072 & 0.071 & 0.021 & 0.074 & 0.023 & 0.064 & 0.014 \\ 
     & & \multicolumn{12}{c}{$\bm{\alpha}=(-0.5,-0.5,0.5,0.5)^\top$, $\sigma_k^2=1,~k=1,\ldots,4$}\\
0 & 0.606 & 0.115 & 0.733 & 0.065 & 0.679 & 0.056 & 0.759 & 0.040 & 0.284 & 0.119 & 0.079 & 0.045 & 0.062 & 0.010 & 0.058 & 0.009 \\ 
0.25 & 0.668 & 0.114 & 0.773 & 0.048 & 0.728 & 0.052 & 0.788 & 0.049 & 0.228 & 0.119 & 0.066 & 0.013 & 0.066 & 0.010 & 0.060 & 0.009 \\ 
0.5 & 0.725 & 0.106 & 0.804 & 0.056 & 0.761 & 0.058 & 0.805 & 0.062 & 0.173 & 0.111 & 0.058 & 0.012 & 0.060 & 0.011 & 0.053 & 0.010 \\ 
0.75 & 0.771 & 0.106 & 0.810 & 0.065 & 0.771 & 0.075 & 0.810 & 0.066 & 0.137 & 0.104 & 0.074 & 0.022 & 0.075 & 0.024 & 0.064 & 0.014 \\ 
 & & \multicolumn{12}{c}{$\bm{\alpha}=(-0.5,-0.5,0.5,0.5)^\top$, $\sigma_k^2=2,~k=1,\ldots,4$}\\
0 & 0.560 & 0.138 & 0.725 & 0.078 & 0.643 & 0.061 & 0.759 & 0.040 & 0.358 & 0.145 & 0.096 & 0.067 & 0.063 & 0.010 & 0.058 & 0.009 \\ 
0.25 & 0.610 & 0.139 & 0.767 & 0.056 & 0.695 & 0.054 & 0.788 & 0.049 & 0.309 & 0.145 & 0.066 & 0.011 & 0.067 & 0.010 & 0.060 & 0.009 \\ 
0.5 & 0.669 & 0.140 & 0.800 & 0.054 & 0.732 & 0.060 & 0.805 & 0.062 & 0.246 & 0.147 & 0.060 & 0.013 & 0.061 & 0.012 & 0.053 & 0.010 \\ 
0.75 & 0.720 & 0.154 & 0.809 & 0.065 & 0.747 & 0.079 & 0.810 & 0.066 & 0.200 & 0.160 & 0.076 & 0.026 & 0.075 & 0.024 & 0.064 & 0.014 \\ \hline
\end{tabular}
\end{table}
\end{landscape}

\begin{landscape}
\begin{table}
 \centering
 \footnotesize
 \caption{\label{sum.4} Results of the simulation study including root mean square error (RMSE), relative bias (RB), and coverage rate (CR) for the estimates of the association parameters between event risk and the current value of the four longitudinal markers. The study is based on 100 simulated datasets, each with a sample size of 500, and $(\rho^\dagger,\rho^*)' = (0.4, 0.5)'$.}
\centering
\begin{tabular}{c|ccc|ccc|ccc|ccc|ccc}
  \hline
 &\multicolumn{3}{c|}{MMJM } &\multicolumn{3}{c|}{TSJM} &\multicolumn{3}{c|}{MTS}&\multicolumn{3}{c|}{MB}&\multicolumn{3}{c}{TRUE } \\    \hline
&  RMSE & RB & CR & RMSE & RB & CR & RMSE & RB & CR & RMSE & RB & CR & RMSE & RB & CR \\ 
  \hline
   & & \multicolumn{12}{c}{$\bm{\alpha}=(-0.2,-0.2,0.2,0.2)^\top$, $\sigma_k^2=0.5,~k=1,\ldots,4$}\\
$\alpha_1$ & 0.108 & 0.113 & 0.900 & 0.113 & 0.159 & 0.860 & 0.096 & 0.030 & 0.890  & 0.119 & 0.021 & 0.940 & 0.093 & 0.141 & 0.910 \\ 
$\alpha_2$ & 0.092 & -0.064 & 0.980 & 0.094 & -0.016 & 0.960 & 0.087 & -0.120 & 0.980 & 0.103 & -0.130 & 0.988 & 0.087 & -0.023 & 0.960 \\ 
$\alpha_3$ & 0.088 & 0.002 & 0.930 & 0.091 & 0.048 & 0.930 & 0.081 & -0.065 & 0.970  & 0.091 & -0.078 & 0.976 & 0.077 & 0.057 & 0.950 \\ 
$\alpha_4$ & 0.085 & 0.041 & 0.940 & 0.090 & 0.082 & 0.920 & 0.079 & -0.039 & 0.950 & 0.084 & -0.068 & 1.000 & 0.077 & 0.057 & 0.940 \\ 
    & & \multicolumn{12}{c}{$\bm{\alpha}=(-0.2,-0.2,0.2,0.2)^\top$, $\sigma_k^2=1,~k=1,\ldots,4$}\\
$\alpha_1$  & 0.119 & 0.119 & 0.890 & 0.123 & 0.153 & 0.890 & 0.097 & -0.054 & 0.900  & 0.132 & -0.025 & 0.940 & 0.093 & 0.141 & 0.910 \\ 
$\alpha_2$  & 0.099 & -0.046 & 0.960 & 0.099 & -0.020 & 0.940 & 0.090 & -0.191 & 0.980 & 0.113 & -0.168 & 1.000  & 0.087 & -0.023 & 0.960 \\ 
$\alpha_3$  & 0.096 & 0.003 & 0.940 & 0.097 & 0.034 & 0.920 & 0.085 & -0.150 & 0.960 & 0.104 & -0.128 & 0.976 & 0.077 & 0.057 & 0.950 \\ 
$\alpha_4$   & 0.095 & 0.063 & 0.960 & 0.099 & 0.081 & 0.920 & 0.082 & -0.112 & 0.960 & 0.095 & -0.114 & 1.000 & 0.077 & 0.057 & 0.940 \\ 
 & & \multicolumn{12}{c}{$\bm{\alpha}=(-0.2,-0.2,0.2,0.2)^\top$, $\sigma_k^2=2,~k=1,\ldots,4$}\\
$\alpha_1$ & 0.142 & 0.148 & 0.880 & 0.138 & 0.158 & 0.880 & 0.102 & -0.168 & 0.880  & 0.149 & -0.091 & 0.952 & 0.093 & 0.141 & 0.910 \\ 
$\alpha_2$ & 0.113 & -0.036 & 0.960 & 0.109 & -0.024 & 0.920 & 0.097 & -0.286 & 0.980 & 0.128 & -0.227 & 0.964 & 0.087 & -0.023 & 0.960 \\ 
$\alpha_3$ & 0.108 & 0.013 & 0.950 & 0.108 & 0.022 & 0.920 & 0.094 & -0.261 & 0.940 & 0.123 & -0.203 & 0.940 & 0.077 & 0.057 & 0.950 \\ 
$\alpha_4$  & 0.113 & 0.083 & 0.940 & 0.112 & 0.087 & 0.890 & 0.090 & -0.213 & 0.960 & 0.113 & -0.184 & 0.988 & 0.077 & 0.057 & 0.940 \\ 
    & & \multicolumn{12}{c}{$\bm{\alpha}=(-0.5,-0.5,0.5,0.5)^\top$, $\sigma_k^2=0.5,~k=1,\ldots,4$}\\
$\alpha_1$ & 0.103 & 0.024 & 0.960 & 0.099 & 0.020 & 0.940 & 0.098 & -0.085 & 0.930 & 0.124 & -0.114 & 0.975 & 0.087 & 0.043 & 0.960 \\ 
$\alpha_2$   & 0.112 & 0.012 & 0.930 & 0.107 & 0.012 & 0.930 & 0.106 & -0.092 & 0.900  & 0.128 & -0.069 & 0.962 & 0.089 & 0.014 & 0.980 \\ 
$\alpha_3$ & 0.095 & 0.044 & 0.950 & 0.092 & 0.036 & 0.950 & 0.088 & -0.071 & 0.970 & 0.137 & -0.092 & 0.937  & 0.083 & 0.050 & 0.950 \\ 
$\alpha_4$ & 0.102 & 0.018 & 0.950 & 0.100 & 0.021 & 0.940 & 0.099 & -0.085 & 0.940 & 0.117 & -0.101 & 0.937  & 0.085 & 0.042 & 0.970 \\ 
     & & \multicolumn{12}{c}{$\bm{\alpha}=(-0.5,-0.5,0.5,0.5)^\top$, $\sigma_k^2=1,~k=1,\ldots,4$}\\
$\alpha_1$  & 0.120 & 0.051 & 0.960 & 0.104 & -0.007 & 0.930 & 0.123 & -0.172 & 0.870  & 0.124 & -0.114 & 0.975 & 0.087 & 0.043 & 0.960 \\ 
$\alpha_2$ & 0.135 & 0.042 & 0.920 & 0.116 & -0.010 & 0.910 & 0.130 & -0.175 & 0.860  & 0.128 & -0.069 & 0.962  & 0.089 & 0.014 & 0.980 \\
$\alpha_3$ & 0.114 & 0.074 & 0.930 & 0.096 & 0.013 & 0.930 & 0.112 & -0.157 & 0.900 & 0.137 & -0.092 & 0.937 & 0.083 & 0.050 & 0.950 \\ 
$\alpha_4$ & 0.122 & 0.039 & 0.920 & 0.106 & -0.011 & 0.920 & 0.124 & -0.176 & 0.890  & 0.117 & -0.101 & 0.937 & 0.085 & 0.042 & 0.970 \\ 
 & & \multicolumn{12}{c}{$\bm{\alpha}=(-0.5,-0.5,0.5,0.5)^\top$, $\sigma_k^2=2,~k=1,\ldots,4$}\\
$\alpha_1$ & 0.160 & 0.098 & 0.900 & 0.116 & -0.041 & 0.930 & 0.168 & -0.287 & 0.710 & 0.214 & -0.322 & 0.833 & 0.087 & 0.043 & 0.960 \\ 
$\alpha_2$ & 0.171 & 0.084 & 0.890 & 0.131 & -0.041 & 0.840 & 0.171 & -0.286 & 0.620 & 0.187 & -0.208 & 0.950 & 0.089 & 0.014 & 0.980 \\ 
$\alpha_3$ & 0.156 & 0.124 & 0.920 & 0.107 & -0.020 & 0.920 & 0.157 & -0.270 & 0.750  & 0.201 & -0.277 & 0.850 & 0.083 & 0.050 & 0.950 \\  
$\alpha_4$ & 0.161 & 0.070 & 0.920 & 0.119 & -0.051 & 0.910 & 0.171 & -0.296 & 0.630  & 0.182 & -0.265 & 0.883 & 0.085 & 0.042 & 0.970 \\ 
   \hline
\end{tabular}
\end{table}
\end{landscape}

\begin{landscape}
\begin{table}
 \centering
 \footnotesize
 \caption{\label{auc.4} Results of the simulation study including AUC and BS for 100 simulated datasets, each with a sample size of 500 and $(\rho^\dagger,\rho^*)' = (0.4, 0.5)'$.}
\centering
\begin{tabular}{c|cccccccc|cccccccc}
  \hline
 &\multicolumn{8}{c|}{AUC} &\multicolumn{8}{c}{BS } \\    \hline
  &\multicolumn{2}{c}{MMJM } &\multicolumn{2}{c}{TSJM} &\multicolumn{2}{c}{MTS}&\multicolumn{2}{c|}{TRUE } & \multicolumn{2}{c}{MMJM } &\multicolumn{2}{c}{TSJM} &\multicolumn{2}{c}{MTS}&\multicolumn{2}{c}{TRUE } \\    \hline
& mean & sd & mean & sd & mean & sd & mean & sd & mean & sd & mean & sd & mean & sd & mean & sd  \\   \hline
  & & \multicolumn{12}{c}{$\bm{\alpha}=(-0.2,-0.2,0.2,0.2)^\top$, $\sigma_k^2=0.5,~k=1,\ldots,4$}\\
0 & 0.647 & 0.125 & 0.682 & 0.054 & 0.598 & 0.054 & 0.719 & 0.043 & 0.240 & 0.137 & 0.062 & 0.012 & 0.061 & 0.009 & 0.058 & 0.008 \\ 
0.25 & 0.661 & 0.122 & 0.707 & 0.057 & 0.639 & 0.060 & 0.721 & 0.049 & 0.224 & 0.138 & 0.068 & 0.013 & 0.068 & 0.011 & 0.064 & 0.010 \\ 
0.5 & 0.693 & 0.128 & 0.729 & 0.065 & 0.668 & 0.072 & 0.739 & 0.062 & 0.191 & 0.139 & 0.063 & 0.014 & 0.063 & 0.012 & 0.058 & 0.011 \\ 
0.75 & 0.735 & 0.130 & 0.743 & 0.065 & 0.688 & 0.071 & 0.756 & 0.067 & 0.162 & 0.139 & 0.079 & 0.026 & 0.075 & 0.018 & 0.069 & 0.014 \\ 
    & & \multicolumn{12}{c}{$\bm{\alpha}=(-0.2,-0.2,0.2,0.2)^\top$, $\sigma_k^2=1,~k=1,\ldots,4$}\\
0 & 0.620 & 0.054 & 0.580 & 0.053 & 0.548 & 0.051 & 0.586 & 0.053 & 0.062 & 0.010 & 0.062 & 0.010 & 0.062 & 0.010 & 0.062 & 0.010 \\ 
0.25 & 0.615 & 0.062 & 0.589 & 0.066 & 0.566 & 0.064 & 0.589 & 0.067 & 0.069 & 0.012 & 0.069 & 0.012 & 0.069 & 0.012 & 0.069 & 0.012 \\ 
0.5 & 0.615 & 0.066 & 0.599 & 0.070 & 0.576 & 0.071 & 0.597 & 0.073 & 0.063 & 0.013 & 0.063 & 0.013 & 0.063 & 0.013 & 0.063 & 0.013 \\ 
0.75 & 0.635 & 0.067 & 0.617 & 0.065 & 0.596 & 0.066 & 0.614 & 0.081 & 0.067 & 0.016 & 0.071 & 0.017 & 0.070 & 0.016 & 0.069 & 0.016 \\ 
   & & \multicolumn{12}{c}{$\bm{\alpha}=(-0.2,-0.2,0.2,0.2)^\top$, $\sigma_k^2=2,~k=1,\ldots,4$}\\
0 & 0.646 & 0.056 & 0.586 & 0.052 & 0.536 & 0.052 & 0.586 & 0.053 & 0.063 & 0.014 & 0.062 & 0.010 & 0.062 & 0.010 & 0.062 & 0.010 \\ 
0.25 & 0.636 & 0.065 & 0.595 & 0.065 & 0.553 & 0.065 & 0.589 & 0.067 & 0.069 & 0.014 & 0.069 & 0.012 & 0.069 & 0.012 & 0.069 & 0.012 \\ 
0.5 & 0.628 & 0.068 & 0.599 & 0.073 & 0.562 & 0.071 & 0.597 & 0.073 & 0.063 & 0.013 & 0.064 & 0.013 & 0.064 & 0.013 & 0.063 & 0.013 \\ 
0.75 & 0.647 & 0.067 & 0.615 & 0.065 & 0.584 & 0.065 & 0.614 & 0.081 & 0.067 & 0.016 & 0.071 & 0.017 & 0.070 & 0.016 & 0.069 & 0.016 \\ 
    & & \multicolumn{12}{c}{$\bm{\alpha}=(-0.5,-0.5,0.5,0.5)^\top$, $\sigma_k^2=0.5,~k=1,\ldots,4$}\\
0 & 0.662 & 0.090 & 0.688 & 0.051 & 0.656 & 0.052 & 0.719 & 0.043 & 0.156 & 0.087 & 0.060 & 0.009 & 0.061 & 0.009 & 0.058 & 0.008 \\ 
0.25 & 0.687 & 0.079 & 0.713 & 0.055 & 0.688 & 0.056 & 0.721 & 0.049 & 0.141 & 0.089 & 0.067 & 0.011 & 0.067 & 0.011 & 0.064 & 0.010 \\ 
0.5 & 0.727 & 0.081 & 0.734 & 0.065 & 0.710 & 0.065 & 0.739 & 0.062 & 0.111 & 0.077 & 0.062 & 0.012 & 0.062 & 0.012 & 0.058 & 0.011 \\ 
0.75 & 0.767 & 0.068 & 0.754 & 0.061 & 0.732 & 0.064 & 0.756 & 0.067 & 0.092 & 0.057 & 0.076 & 0.020 & 0.075 & 0.018 & 0.069 & 0.014 \\ 
     & & \multicolumn{12}{c}{$\bm{\alpha}=(-0.5,-0.5,0.5,0.5)^\top$, $\sigma_k^2=1,~k=1,\ldots,4$}\\
0 & 0.650 & 0.106 & 0.683 & 0.053 & 0.628 & 0.054 & 0.719 & 0.043 & 0.202 & 0.110 & 0.060 & 0.009 & 0.061 & 0.009 & 0.058 & 0.008 \\ 
0.25 & 0.677 & 0.096 & 0.708 & 0.056 & 0.667 & 0.058 & 0.721 & 0.049 & 0.177 & 0.110 & 0.067 & 0.011 & 0.068 & 0.011 & 0.064 & 0.010 \\ 
0.5 & 0.716 & 0.100 & 0.730 & 0.065 & 0.693 & 0.068 & 0.739 & 0.062 & 0.141 & 0.104 & 0.062 & 0.012 & 0.062 & 0.012 & 0.058 & 0.011 \\ 
0.75 & 0.756 & 0.088 & 0.750 & 0.063 & 0.714 & 0.067 & 0.756 & 0.067 & 0.117 & 0.091 & 0.076 & 0.019 & 0.075 & 0.018 & 0.069 & 0.014 \\ 
 & & \multicolumn{12}{c}{$\bm{\alpha}=(-0.5,-0.5,0.5,0.5)^\top$, $\sigma_k^2=2,~k=1,\ldots,4$}\\
0 & 0.647 & 0.125 & 0.682 & 0.054 & 0.598 & 0.054 & 0.719 & 0.043 & 0.240 & 0.137 & 0.062 & 0.012 & 0.061 & 0.009 & 0.058 & 0.008 \\ 
0.25 & 0.661 & 0.122 & 0.707 & 0.057 & 0.639 & 0.060 & 0.721 & 0.049 & 0.224 & 0.138 & 0.068 & 0.013 & 0.068 & 0.011 & 0.064 & 0.010 \\ 
0.5 & 0.693 & 0.128 & 0.729 & 0.065 & 0.668 & 0.072 & 0.739 & 0.062 & 0.191 & 0.139 & 0.063 & 0.014 & 0.063 & 0.012 & 0.058 & 0.011 \\ 
0.75 & 0.735 & 0.130 & 0.743 & 0.065 & 0.688 & 0.071 & 0.756 & 0.067 & 0.162 & 0.139 & 0.079 & 0.026 & 0.075 & 0.018 & 0.069 & 0.014 \\ 
  \hline
\end{tabular}
\end{table}
\end{landscape}

\begin{landscape}
\begin{table}
 \centering
 \footnotesize
 \caption{\label{sum.6} Results of the simulation study including root mean square error (RMSE), relative bias (RB), and coverage rate (CR) for the estimates of the association parameters between event risk and the current value of the four longitudinal markers. The study is based on 100 simulated datasets, each with a sample size of 500, and $(\rho^\dagger,\rho^*)' = (0.1, 0.5)'$.}
\centering
\begin{tabular}{c|ccc|ccc|ccc|ccc|ccc}
  \hline
 &\multicolumn{3}{c|}{MMJM } &\multicolumn{3}{c|}{TSJM} &\multicolumn{3}{c|}{MTS}&\multicolumn{3}{c|}{MB}&\multicolumn{3}{c}{TRUE } \\    \hline
&  RMSE & RB & CR & RMSE & RB & CR  & RMSE & RB & CR & RMSE & RB & CR & RMSE & RB & CR \\ 
  \hline
   & & \multicolumn{12}{c}{$\bm{\alpha}=(-0.2,-0.2,0.2,0.2)^\top$, $\sigma_k^2=0.5,~k=1,\ldots,4$}\\
$\alpha_1$ & 0.094 & 0.007 & 0.970 & 0.097 & 0.034 & 0.960 & 0.088 & -0.062 & 0.970  & 0.114 & -0.124 & 0.964 & 0.081 & 0.049 & 0.980 \\ 
$\alpha_2$  & 0.105 & -0.007 & 0.940 & 0.106 & 0.005 & 0.920 & 0.097 & -0.091 & 0.950   & 0.108 & -0.110 & 0.964 & 0.082 & 0.039 & 0.970 \\
$\alpha_3$  & 0.096 & -0.052 & 0.940 & 0.098 & -0.025 & 0.960 & 0.091 & -0.116 & 0.960  & 0.110 & -0.169 & 0.964 & 0.088 & -0.003 & 0.960 \\ 
$\alpha_4$  & 0.102 & 0.059 & 0.950 & 0.102 & 0.071 & 0.950 & 0.091 & -0.030 & 0.970  & 0.117 & -0.037 & 0.976 & 0.093 & 0.088 & 0.960 \\ 
    & & \multicolumn{12}{c}{$\bm{\alpha}=(-0.2,-0.2,0.2,0.2)^\top$, $\sigma_k^2=1,~k=1,\ldots,4$}\\
$\alpha_1$ & 0.106 & 0.010 & 0.960 & 0.105 & 0.005 & 0.940 & 0.093 & -0.150 & 0.970  & 0.135 & -0.208 & 0.962 & 0.081 & 0.049 & 0.980 \\ 
$\alpha_2$  & 0.126 & 0.018 & 0.910 & 0.117 & -0.020 & 0.910 & 0.103 & -0.175 & 0.940  & 0.124 & -0.162 & 0.962 & 0.082 & 0.039 & 0.970 \\ 
$\alpha_3$ & 0.104 & -0.032 & 0.950 & 0.103 & -0.047 & 0.940 & 0.095 & -0.196 & 0.940  & 0.126 & -0.272 & 0.962 & 0.088 & -0.003 & 0.960 \\ 
$\alpha_4$  & 0.114 & 0.071 & 0.950 & 0.107 & 0.044 & 0.930 & 0.093 & -0.120 & 0.960 & 0.130 & -0.064 & 0.975 & 0.093 & 0.088 & 0.960 \\ 
   & & \multicolumn{12}{c}{$\bm{\alpha}=(-0.2,-0.2,0.2,0.2)^\top$, $\sigma_k^2=2,~k=1,\ldots,4$}\\
$\alpha_1$ & 0.127 & 0.043 & 0.950 & 0.117 & -0.036 & 0.920 & 0.103 & -0.271 & 0.920 & 0.164 & -0.283 & 0.952 & 0.081 & 0.049 & 0.980 \\  
$\alpha_2$ & 0.158 & 0.068 & 0.900 & 0.134 & -0.054 & 0.890 & 0.114 & -0.285 & 0.910  & 0.153 & -0.278 & 0.921 & 0.082 & 0.039 & 0.970 \\ 
$\alpha_3$  & 0.119 & 0.007 & 0.950 & 0.112 & -0.072 & 0.930 & 0.103 & -0.302 & 0.940   & 0.149 & -0.402 & 0.921 & 0.088 & -0.003 & 0.960 \\
$\alpha_4$  & 0.144 & 0.119 & 0.930 & 0.115 & 0.002 & 0.930 & 0.100 & -0.241 & 0.950  & 0.158 & -0.136 & 0.952 & 0.093 & 0.088 & 0.960 \\ 
    & & \multicolumn{12}{c}{$\bm{\alpha}=(-0.5,-0.5,0.5,0.5)^\top$, $\sigma_k^2=0.5,~k=1,\ldots,4$}\\
$\alpha_1$ & 0.114 & 0.048 & 0.930 & 0.102 & 0.006 & 0.940 & 0.101 & -0.080 & 0.940 & 0.117 & -0.094 & 0.948 & 0.099 & 0.060 & 0.920 \\ 
$\alpha_2$ & 0.121 & 0.026 & 0.920 & 0.111 & 0.004 & 0.920 & 0.111 & -0.085 & 0.920  & 0.123 & -0.101 & 0.935 & 0.103 & 0.038 & 0.900 \\ 
$\alpha_3$  & 0.114 & 0.043 & 0.940 & 0.106 & 0.000 & 0.950 & 0.106 & -0.089 & 0.910 & 0.101 & -0.093 & 0.974  & 0.099 & 0.048 & 0.970 \\ 
$\alpha_4$ & 0.111 & 0.025 & 0.920 & 0.104 & -0.003 & 0.930 & 0.106 & -0.090 & 0.920 & 0.112 & -0.071 & 0.961 & 0.099 & 0.035 & 0.900 \\ 
     & & \multicolumn{12}{c}{$\bm{\alpha}=(-0.5,-0.5,0.5,0.5)^\top$, $\sigma_k^2=1,~k=1,\ldots,4$}\\
$\alpha_1$ & 0.139 & 0.071 & 0.940 & 0.110 & -0.043 & 0.910 & 0.128 & -0.178 & 0.840 & 0.143 & -0.161 & 0.974 & 0.099 & 0.060 & 0.920 \\ 
$\alpha_2$ & 0.141 & 0.056 & 0.930 & 0.118 & -0.044 & 0.920 & 0.136 & -0.184 & 0.870  & 0.150 & -0.161 & 0.922 & 0.103 & 0.038 & 0.900 \\  
$\alpha_3$ & 0.137 & 0.069 & 0.910 & 0.113 & -0.049 & 0.900 & 0.133 & -0.189 & 0.840 & 0.131 & -0.159 & 0.961  & 0.099 & 0.048 & 0.970 \\ 
$\alpha_4$ & 0.131 & 0.051 & 0.940 & 0.112 & -0.051 & 0.910 & 0.133 & -0.186 & 0.850 & 0.135 & -0.137 & 0.922 & 0.099 & 0.035 & 0.900 \\ 
 & & \multicolumn{12}{c}{$\bm{\alpha}=(-0.5,-0.5,0.5,0.5)^\top$, $\sigma_k^2=2,~k=1,\ldots,4$}\\
$\alpha_1$ & 0.197 & 0.144 & 0.920 & 0.129 & -0.100 & 0.830 & 0.176 & -0.303 & 0.660 & 0.184 & -0.251 & 0.961 & 0.099 & 0.060 & 0.920 \\ 
$\alpha_2$  & 0.183 & 0.104 & 0.940 & 0.135 & -0.105 & 0.850 & 0.183 & -0.312 & 0.600 & 0.189 & -0.243 & 0.896 & 0.103 & 0.038 & 0.900 \\ 
$\alpha_3$  & 0.180 & 0.129 & 0.900 & 0.132 & -0.110 & 0.860 & 0.182 & -0.317 & 0.640 & 0.176 & -0.251 & 0.948  & 0.099 & 0.048 & 0.970 \\ 
$\alpha_4$ & 0.181 & 0.116 & 0.930 & 0.131 & -0.110 & 0.870 & 0.181 & -0.311 & 0.620 & 0.173 & -0.228 & 0.935 & 0.099 & 0.035 & 0.900 \\ 
  \hline
\end{tabular}
\end{table}
\end{landscape}

\begin{landscape}
\begin{table}
 \centering
 \footnotesize
 \caption{\label{auc.6} Results of the simulation study including AUC and BS for 100 simulated datasets, each with a sample size of 500 and $(\rho^\dagger,\rho^*)' = (0.6,0.8)'$.}
\centering
\begin{tabular}{c|cccccccc|cccccccc}
  \hline
 &\multicolumn{8}{c|}{AUC} &\multicolumn{8}{c}{BS } \\    \hline
  &\multicolumn{2}{c}{MMJM } &\multicolumn{2}{c}{TSJM} &\multicolumn{2}{c}{MTS}&\multicolumn{2}{c|}{TRUE } & \multicolumn{2}{c}{MMJM } &\multicolumn{2}{c}{TSJM} &\multicolumn{2}{c}{MTS}&\multicolumn{2}{c}{TRUE } \\    \hline
& mean & sd & mean & sd & mean & sd & mean & sd & mean & sd & mean & sd & mean & sd & mean & sd  \\   \hline
  & & \multicolumn{12}{c}{$\bm{\alpha}=(-0.2,-0.2,0.2,0.2)^\top$, $\sigma_k^2=0.5,~k=1,\ldots,4$}\\
0 & 0.595 & 0.056 & 0.562 & 0.057 & 0.547 & 0.056 & 0.574 & 0.052 & 0.061 & 0.009 & 0.061 & 0.009 & 0.061 & 0.010 & 0.061 & 0.009 \\ 
0.25 & 0.594 & 0.065 & 0.575 & 0.065 & 0.565 & 0.063 & 0.586 & 0.065 & 0.068 & 0.011 & 0.069 & 0.011 & 0.069 & 0.011 & 0.068 & 0.011 \\ 
0.5 & 0.613 & 0.076 & 0.598 & 0.080 & 0.589 & 0.078 & 0.604 & 0.083 & 0.062 & 0.012 & 0.062 & 0.012 & 0.062 & 0.012 & 0.062 & 0.011 \\ 
0.75 & 0.615 & 0.077 & 0.600 & 0.077 & 0.592 & 0.075 & 0.603 & 0.073 & 0.068 & 0.014 & 0.071 & 0.014 & 0.070 & 0.014 & 0.070 & 0.014 \\ 
    & & \multicolumn{12}{c}{$\bm{\alpha}=(-0.2,-0.2,0.2,0.2)^\top$, $\sigma_k^2=1,~k=1,\ldots,4$}\\
0 & 0.611 & 0.060 & 0.560 & 0.055 & 0.537 & 0.056 & 0.574 & 0.052 & 0.061 & 0.009 & 0.061 & 0.009 & 0.062 & 0.010 & 0.061 & 0.009 \\ 
0.25 & 0.607 & 0.066 & 0.574 & 0.062 & 0.555 & 0.064 & 0.586 & 0.065 & 0.068 & 0.011 & 0.069 & 0.011 & 0.069 & 0.011 & 0.068 & 0.011 \\ 
0.5 & 0.623 & 0.078 & 0.599 & 0.082 & 0.580 & 0.081 & 0.604 & 0.083 & 0.062 & 0.012 & 0.062 & 0.012 & 0.062 & 0.012 & 0.062 & 0.011 \\ 
0.75 & 0.620 & 0.076 & 0.597 & 0.075 & 0.582 & 0.075 & 0.603 & 0.073 & 0.068 & 0.014 & 0.071 & 0.014 & 0.070 & 0.014 & 0.070 & 0.014 \\ 
   & & \multicolumn{12}{c}{$\bm{\alpha}=(-0.2,-0.2,0.2,0.2)^\top$, $\sigma_k^2=2,~k=1,\ldots,4$}\\
0 & 0.638 & 0.066 & 0.564 & 0.058 & 0.526 & 0.056 & 0.574 & 0.052 & 0.062 & 0.011 & 0.062 & 0.009 & 0.062 & 0.010 & 0.061 & 0.009 \\ 
0.25 & 0.629 & 0.071 & 0.576 & 0.065 & 0.543 & 0.064 & 0.586 & 0.065 & 0.068 & 0.011 & 0.069 & 0.011 & 0.069 & 0.011 & 0.068 & 0.011 \\ 
0.5 & 0.645 & 0.080 & 0.598 & 0.086 & 0.569 & 0.085 & 0.604 & 0.083 & 0.062 & 0.012 & 0.062 & 0.012 & 0.062 & 0.012 & 0.062 & 0.011 \\ 
0.75 & 0.639 & 0.076 & 0.591 & 0.073 & 0.568 & 0.074 & 0.603 & 0.073 & 0.067 & 0.014 & 0.072 & 0.014 & 0.070 & 0.014 & 0.070 & 0.014 \\ 
    & & \multicolumn{12}{c}{$\bm{\alpha}=(-0.5,-0.5,0.5,0.5)^\top$, $\sigma_k^2=0.5,~k=1,\ldots,4$}\\
0 & 0.661 & 0.077 & 0.656 & 0.051 & 0.627 & 0.049 & 0.688 & 0.049 & 0.146 & 0.085 & 0.062 & 0.009 & 0.063 & 0.010 & 0.060 & 0.009 \\ 
0.25 & 0.699 & 0.067 & 0.688 & 0.060 & 0.668 & 0.063 & 0.706 & 0.051 & 0.116 & 0.065 & 0.069 & 0.011 & 0.069 & 0.011 & 0.066 & 0.011 \\ 
0.5 & 0.735 & 0.064 & 0.715 & 0.067 & 0.698 & 0.069 & 0.737 & 0.064 & 0.082 & 0.046 & 0.060 & 0.013 & 0.061 & 0.013 & 0.057 & 0.012 \\ 
0.75 & 0.770 & 0.056 & 0.741 & 0.055 & 0.725 & 0.058 & 0.752 & 0.061 & 0.075 & 0.033 & 0.076 & 0.017 & 0.076 & 0.017 & 0.068 & 0.013 \\ 
     & & \multicolumn{12}{c}{$\bm{\alpha}=(-0.5,-0.5,0.5,0.5)^\top$, $\sigma_k^2=1,~k=1,\ldots,4$}\\
0 & 0.651 & 0.095 & 0.648 & 0.051 & 0.602 & 0.049 & 0.688 & 0.049 & 0.188 & 0.109 & 0.062 & 0.010 & 0.063 & 0.010 & 0.060 & 0.009 \\ 
0.25 & 0.681 & 0.091 & 0.679 & 0.063 & 0.644 & 0.065 & 0.706 & 0.051 & 0.161 & 0.100 & 0.070 & 0.012 & 0.070 & 0.012 & 0.066 & 0.011 \\ 
0.5 & 0.725 & 0.083 & 0.705 & 0.067 & 0.674 & 0.073 & 0.737 & 0.064 & 0.114 & 0.080 & 0.062 & 0.014 & 0.061 & 0.013 & 0.057 & 0.012 \\ 
0.75 & 0.772 & 0.069 & 0.733 & 0.057 & 0.706 & 0.058 & 0.752 & 0.061 & 0.093 & 0.067 & 0.078 & 0.022 & 0.076 & 0.017 & 0.068 & 0.013 \\ 
 & & \multicolumn{12}{c}{$\bm{\alpha}=(-0.5,-0.5,0.5,0.5)^\top$, $\sigma_k^2=2,~k=1,\ldots,4$}\\
0 & 0.651 & 0.119 & 0.648 & 0.054 & 0.577 & 0.050 & 0.688 & 0.049 & 0.224 & 0.131 & 0.062 & 0.010 & 0.063 & 0.010 & 0.060 & 0.009 \\ 
0.25 & 0.661 & 0.120 & 0.669 & 0.060 & 0.616 & 0.066 & 0.706 & 0.051 & 0.217 & 0.137 & 0.069 & 0.011 & 0.070 & 0.012 & 0.066 & 0.011 \\ 
0.5 & 0.692 & 0.125 & 0.691 & 0.074 & 0.643 & 0.078 & 0.737 & 0.064 & 0.182 & 0.137 & 0.061 & 0.013 & 0.062 & 0.013 & 0.057 & 0.012 \\ 
0.75 & 0.753 & 0.113 & 0.723 & 0.059 & 0.679 & 0.059 & 0.752 & 0.061 & 0.145 & 0.126 & 0.077 & 0.017 & 0.075 & 0.016 & 0.068 & 0.013 \\ 
   \hline
\end{tabular}
\end{table}
\end{landscape}

\section{Applications}
\subsection{Application 1: PBC2 data }
In this section, we applied the TSJM and the MMJM approaches to analyze the PBC2 data  \cite{lin2002modeling} and predict the risk of death.  The dataset includes 312 patients
with primary biliary cirrhosis
who were enrolled in clinical trials at the Mayo Clinic from 1974 to 1984. The average follow-up time was 8.19 years, with a median of 5 visits and a maximum of 16 visits. In this analysis, the event of interest is death without transplantation, while subjects who are alive at the end of the study or have been transplanted are considered right-censored ($44.87\%$ died). The complete data can be found in many R-packages, such as \texttt{joineRML}, which provides a description at the following URL: \url{https://rdrr.io/cran/joineRML/man/pbc2.html}. We consider five Gaussian biological markers, including albumin (in mg/dL), logarithm transformation of alkaline (alkaline phosphatase (in U/L), logarithm transformation of SGOT (in U/mL), logarithm transformation of platelets (platelets per cubic mL/1000), and logarithm transformation of serum bilirubin (serum bilirubin in mg/dl).\\
The sub-model for the time to death depends on the current value of the included markers and is adjusted for the treatment group (drug) and the patient's standardized age at enrollment. The change over time of the markers is described using linear mixed models, which include a linear time trend with random intercepts and random slopes. \\
The estimated parameters for MMJM and TSJM are provided in Table B.1 of the Appendix B. This table shows that the differences in the estimated parameters for the longitudinal sub-models and the association parameters are negligible. \\
Table \ref{aucestpbc} compares the predictive abilities of the TSJM and the MMJM, as measured by the AUC and BS, at the landmark times $s=0,2,4,6,8$ with prediction window of two years. We randomly split the data into a $70\%$ learning sample and a $30\%$ validation sample. Based on the results, although the values of AUC for the MMJM are  larger than those for TSJM for $s= 0,2$ and $4$, the differences are not statistically significant based on the test given by the \texttt{timeROC} package \cite{blanche2019package}. This table also shows that the values of BS for the MMJM are smaller than those for the TSJM, but the differences are not statistically significant.

\begin{table}
\footnotesize
 \centering
 \caption{AUC and Brier score computed for landmark times $s=0,2,4,6,8$ years and prediction window of two years for the comparison between the TSJM and the MMJM for PBC2 data. \label{aucestpbc}  }
\centering
\begin{tabular}{c|ccccc|ccccc}
  \hline
  &\multicolumn{5}{c}{AUC } &\multicolumn{5}{c}{BS } \\ \hline
    &\multicolumn{2}{c}{TSJM  }&\multicolumn{2}{c}{MMJM}& Comparison &\multicolumn{2}{c}{TSJM  }&\multicolumn{2}{c}{MMJM}& Comparison\\ \hline
  & Est. & SD. & Est. & SD. & p-value  & Est. & SD. & Est. & SD. &  p-value \\ 
 \hline
$s=0$  & 0.905 & 0.041 & 0.908 & 0.042 & 0.680 & 0.082 & 0.023 & 0.080 & 0.022  & 0.769 \\ 
$s=2$ & 0.964 & 0.020 & 0.973 & 0.016 & 0.379 & 0.062 & 0.018 & 0.051 & 0.020  & 0.307 \\ 
$s=4$ & 0.898 & 0.046 & 0.925 & 0.039 & 0.228 & 0.121 & 0.035 & 0.097 & 0.034  & 0.295 \\ 
$s=6$ & 0.865 & 0.113 & 0.850 & 0.127 & 0.383 & 0.102 & 0.046 & 0.070 & 0.036 & 0.070 \\ 
$s=8$ & 1.000 & 0.000 & 0.952 & 0.062 & 0.440 & 0.143 & 0.078 & 0.082 & 0.052 & 0.328 \\ 
   \hline
\end{tabular}
\vspace*{30pt}
\end{table}
\subsection{Application 2: 3C study}
\subsubsection{Data description}
In this section, we analyzed a subset of the Three-City (3C) Study \cite{antoniak2003vascular} which is a French cohort study in which participants aged 65 or older were recruited from three cities (Bordeaux, Dijon, and Montpellier) and followed for over 10 years to investigate the association between vascular disease and dementia. The data includes socio-demographic, general health information, and cognitive test scores. The application only includes the center of Dijon and Bordeaux were MRI exams were conducted at years 0, 4 and also 10 years for the center of Bordeaux. Except for the MRI markers, the other longitudinal markers were measured at baseline and at each follow-up visit (years 2, 4, 7, 10, 12, and also 14 and 17 in Bordeaux only).
The participants' dementia was diagnosed in a two-step procedure, where the suspected cases of dementia were examined by clinicians and validated by an independent expert committee of neurologists and geriatricians. In addition to dementia, the exact time of death was collected and is considered as a competing risk.\\
We considered $N = 2133$ subjects who were dementia-free at the beginning of the study and had at least one measurement for each of the longitudinal markers.
Out of these subjects, 230 were diagnosed with incident dementia, and 308 died before developing dementia. Figure C.1 shows the cumulative incidence function for dementia and death for this sample.\\
We considered seventeen longitudinal markers:   three  cardio-metabolic  markers (body mass index (BMI), diastolic blood pressure (DBP), and systolic blood pressure (SBP)), the total number of medications (TOTMED), depressive symptomatology  measured  
using the Center for Epidemiologic Studies-Depression scale (CESDT, the lower the better), functional dependency assessed using Instrumental Activity of Daily Living scale (IADL, the lower the better), four cognitive tests (the visual retention test of Benton (BENTON, number of correct responses among 15), the Isaac set test of verbal fluency (ISA, total number of words given in 4 semantic categories in 15 seconds), the trail making tests A and B (TMTA and TMTB, number of correct moves by minutes); the total intracranial volume (TIV) and four  biomarkers of neurodegeneration, including white matter volume (WMV), gray matter volume (GMV), left hippocampal volume (LHIPP) and right hippocampal volume (RHIPP); two markers of vascular brain lesions including volumes of White Matter Hyperintensities in the periventricular (Peri) and deep (Deep) white matter.
Figure C.2 shows individual trajectories of the longitudinal markers in the 3C study, illustrating time-dependent variables. The data were pre-transformed using splines to satisfy the normality assumption of linear mixed models
\cite{devaux2022random,proust2015estimation}.
The minimum, maximum, and median number of repeated measurements for SBP, DBP, CESDT, BENTON, ISA, TOTMED, and IADL are 1, 8, and 5, respectively. For BMI, TMTA, and TMTB, the minimum, maximum, and median number of repeated measurements are 1, 7, and 4, respectively. The minimum, maximum, and median number of repeated measurements for WMV, GMW, TIV, RHIPP, and LHIPP are 1, 3, and 2, respectively. For Peri and Deep (which were not measured at year 10), there are at most two repeated measures, with a median equal to 2. \\
The analyses were also adjusted for age ($\frac{\rm{age~ at~enrollment-65}}{10}$), education level (Educ, 0=less than 10 years of schooling, 1=at least 10 years of schooling), sex (0=male, 1=female), diabetes status at baseline (1=yes), and APOE4 allele carrier status (the main genetic susceptibility factor for Alzheimer's disease, 1=presence of APOE4). 
\subsubsection{Data analysis}
For analyzing the 3C data using the proposed approach, we consider a cause-specific hazards model \cite{pintilie2006competing,beyersmann2011competing}.
Using this method, one can estimate the failure rate for each of the competing events separately. When dementia is the event of interest,
death without dementia should be treated as censored, along with conventional censored observations. The same procedure can be applied to death when it becomes an event of interest.\\
The longitudinal markers are modeled by a mixed-effects model adjusted for five time-invariant explanatory variables: APOE, sex, diabetes, Educ, and age. As a non linear time trend was previously observed for SBP, DBP, BMI, CESDT, BENTON, ISA, TOTMED, TMTA, TMTB, and IADL \cite{devaux2022random}, a quadratic time-trend is considered for both fixed and random effects, as follows:
\begin{eqnarray}\label{13}
Y_{ik}(t)&=&\eta_{ik}(t|\bm{\beta}_k,\bm{b}_{ik})+\varepsilon_{ikt}\\\nonumber
&=&\beta_{0k}+\beta_{1k}t+\beta_{2k}t^2+\beta_{3k}APOE_{i}+\beta_{4k}Sex_{i} \\\nonumber &+&\beta_{5k}Diabetes_{i}+\beta_{6k}Educ_{i}+\beta_{7k}Age_{i}+
b_{0ki}+b_{1ki} t+b_{1ki} t^2+\varepsilon_{ikt}.\nonumber
\end{eqnarray}
For other longitudinal markers, a linear mixed model is considered:
\begin{eqnarray}\label{12}
Y_{ik}(t)&=&\eta_{ik}(t|\bm{\beta}_k,\bm{b}_{ik})+\varepsilon_{ikt}\\\nonumber
&=&\beta_{0k}+\beta_{1k}t+\beta_{2k}APOE_{i}+\beta_{3k}Sex_{i}\\\nonumber &+& \beta_{4k}Diabetes_{i}+\beta_{5k}Educ_{i}+\beta_{6k}Age_{i}+
b_{0ki}+b_{1ki} t+\varepsilon_{ikt}.\nonumber
\end{eqnarray}
To model dementia and death, we utilize proportional hazard models that are adjusted for the five time-invariant explanatory variables used to model the longitudinal markers. These models are also dependent on the current value of the markers. Let $\lambda_{il}(t)$ denote the cause-specific hazard function for subject $i$, where $l=1$ for dementia and $l=2$ for death. The proportional hazard model takes the following form:
\begin{eqnarray}\label{hh}
\lambda_{il}(t)&=&\lambda_{0l}(t)
\exp(\gamma_{1l}APOE_{i}+\gamma_{2l}Sex_{i}\\\nonumber &+&\gamma_{3l}Diabetes_{i}+ \gamma_{4l}Educ_{i}+\gamma_{5l}Age_{i} +
\sum_{k=1}^{17} \alpha_{kl} \eta_{ik}(t|\bm{\beta}_k,\bm{b}_{ik})).\nonumber
\end{eqnarray}
Tables C.1 to C.4 in the web Appendix C display the estimated regression coefficients from the mixed effects submodels from the one-marker joints models for each marker and either time to dementia or time to death (stage 1 of TSJM). Also, Table \ref{alphas} presents the estimated 
 parameters from the proportional hazard models for dementia and death, respectively (stage 2 of TSJM).
As expected, high values of DBP, TOTMED and IADL and low values of BMI, BENTON, ISA and TMTB are associated with higher risk of dementia. The epsilon 4 allele of APOE also increases the risk. The association of high values of TMTA and VTI as well as high educational level with a higher risk of dementia could appear counter-intuitive, but such results are frequently found after detailed adjustment on current cognitive level. For instance, for a given current cognitive level, subjects with high educational level are those who declined the most.

\begin{table}
 \footnotesize
 \def\~{\hphantom{0}}
 \caption{\label{alphas} Parameter estimates (Est.), standard deviation (SD.), and $95\%$ credible intervals for the association parameters and the regression coefficients of the explanatory variables for dementia and death. }
\centering
\begin{tabular}{c|cccc|cccc}
  \hline
  &\multicolumn{4}{c|}{Dementia } &\multicolumn{4}{c}{Death} \\ \hline
Biomarker  & Est. & SD. & $2.5\%$ & $97.5\%$ & Est. & SD. & $2.5\%$ & $97.5\%$ \\ 
 \hline
SBP & -0.129 & 0.089 & -0.310 & 0.043 & -2.616 & 0.128 & -2.870 & -2.371 \\ 
DBP & 0.222 & 0.086 & 0.053 & 0.394 & 0.398 & 0.077 & 0.252 & 0.551 \\ 
BMI & -0.059 & 0.022 & -0.102 & -0.019 & 0.033 & 0.017 & -0.000 & 0.066 \\ 
CESDT & -0.005 & 0.062 & -0.129 & 0.1160 & -0.269 & 0.065 & -0.399 & -0.142 \\ 
BENTON & -1.198 & 0.145 & -1.479 & -0.888 & -1.466 & 0.092 & -1.647 & -1.292 \\ 
ISA & -0.442 & 0.085 & -0.616 & -0.269 & -0.254 & 0.076 & -0.408 & -0.117 \\ 
TOTMED & 0.119 & 0.062 & 0.002 & 0.246 & 0.359 & 0.052 & 0.254 & 0.459 \\ 
TMTA & 0.431 & 0.082 & 0.264 & 0.582 & 0.031 & 0.069 & -0.111 & 0.164 \\ 
TMTB & -0.803 & 0.139 & -1.068 & -0.548 & -0.257 & 0.083 & -0.426 & -0.101 \\ 
IADL & 0.377 & 0.074 & 0.229 & 0.522 & 0.174 & 0.051 & 0.070 & 0.272 \\ 
WMV & -0.069 & 0.038 & -0.143 & 0.008 & 0.090 & 0.034 & 0.026 & 0.159 \\ 
GMW & -0.009 & 0.029 & -0.068 & 0.044 & 0.017 & 0.027 & -0.037 & 0.069 \\ 
TIV & 0.065 & 0.018 & 0.030 & 0.099 & -0.008 & 0.016 & -0.038 & 0.022 \\ 
RHIPP & 0.002 & 0.031 & -0.062 & 0.064 & -0.083 & 0.024 & -0.130 & -0.037 \\ 
LHIPP & -0.039 & 0.031 & -0.102 & 0.022 & 0.067 & 0.024 & 0.019 & 0.115 \\ 
Peri & 0.019 & 0.013 & -0.007 & 0.042 & -0.036 & 0.016 & -0.066 & -0.006 \\ 
Deep & 0.066 & 0.043 & -0.017 & 0.147 & 0.059 & 0.041 & -0.021 & 0.139 \\ \hline
APOE4 & 0.845 & 0.199 & 0.459 & 1.244 & 0.239 & 0.200 & -0.157 & 0.616 \\ 
Sex & 0.419 & 0.250 & -0.071 & 0.909 & -1.799 & 0.220 & -2.230 & -1.352 \\ 
Diabetes  & 0.221 & 0.262 & -0.313 & 0.710 & 1.235 & 0.213 & 0.829 & 1.658 \\ 
Educ & 0.767 & 0.212 & 0.346 & 1.184 & 0.098 & 0.172 & -0.249 & 0.435 \\ 
Age & 0.198 & 0.259 & -0.315 & 0.696 & 0.475 & 0.242 & -0.019 & 0.946 \\
   \hline
\end{tabular}
\end{table}

\subsubsection{Risk prediction}
For risk prediction, we randomly split the data into a learning sample of 1493 ($70\%$) and a validation sample of 640 ($30\%$). The parameters were estimated by applying TSJM to the training data, and the risk  predictions were calculated for subjects from the validation data set.
With competing risks, the formula \eqref{dp} is not applicable.\\ For this context, let $\Delta_i \in \{1,2\}$ be the event indicator corresponding to the competing risks. Let $\Delta_i=1$ denote the event of interest (that is dementia); the predicted risk of this event between times $s$ and $s+t$ is given by 
\begin{eqnarray*}\label{dpcr}
\pi_i^{C R}(s+t \mid s) =\mathrm{P}\left(s<T_i \leq s+t, \Delta_i=1 \mid T_i>s, \mathcal{Y}_i(s), \boldsymbol{\mathcal { X }}_i(s) ; \boldsymbol{\theta}\right),
\end{eqnarray*}
which is computed by the following  first order approximation:
\begin{eqnarray*}\label{dpcr2}
\hat\pi_i^{CR}(s+t \mid s)  =\frac{\int_{s}^{s+t} 
 \exp\{-\sum_{l=1}^{2} \Lambda_l(u\mid  \hat{\bm{b}}_{il},\bm{\mathcal{X}}_{i}(s); \hat{\bm{\theta}})\}
 \lambda_1(u\mid  \hat{\bm{b}}_{i1}, \bm{\mathcal{X}}_{i}(s); \hat{\bm{\theta}})\}du}{\exp\{-\sum_{l=1}^{2} \Lambda_l(s\mid  \hat{\bm{b}}_{il},\bm{\mathcal{X}}_{i}(s); \hat{\bm{\theta}})\}},
\end{eqnarray*}
where $\hat{\bm{\theta}}$ and $\hat{\bm{b}}_{il}$ are defined as in section \ref{indrisk}. Table \ref{auc3c} shows the values of AUC and BS for the 
validation sample, showing an excellent predictive ability for the 5-year risk of dementia of the model estimated by TSJM.
\begin{table}
 \caption{\label{auc3c} AUC and Brier score computed by validation set  for landmark times $s=0,5,10$ years and prediction window of 5 years for the 3C data.  }
\centering
\begin{tabular}{c|cc|cc}
  \hline
  &\multicolumn{2}{c|}{AUC }&\multicolumn{2}{c}{BS} \\ \hline
  & Est. & SD. & Est. & SD.  \\ 
   \hline
$s=0$ & 0.958 & 0.015 & 0.017 & 0.005\\ 
$s=5$ & 0.940 & 0.011 & 0.191 & 0.018 \\ 
$s=10$ & 0.932 & 0.027 & 0.083 & 0.028 \\ 
  \hline 
\end{tabular}
\end{table}
\section{Discussion}
In this paper, we propose a two-stage approach for joint modeling of multiple longitudinal markers and time-to-event data. This approach is useful when the number of longitudinal markers is large and the estimation of the full joint model including all the markers is not feasible with available softwares. It can also be useful to estimate flexible joint models involving non-linear or time-dependent effects of the markers on the event risk that are not handled by existing software for JM.
We refer to the proposed methods as TSJM, which stands for two-stage joint modeling. 
The first stage of the proposed approach involves the joint modeling of each longitudinal marker and the time-to-event. The second stage involves
fitting a proportional hazard model including predicted functions of the individual marker trajectories obtained at stage 1 as time-dependent covariates.
The computation of dynamic prediction from this TSJM approach is also explained. The simulation study shows that TSJM has performance close to the full joint multi-marker model and better than the standard two-stage approach with respect to parameter estimation and predictive abilities. Two applications to real data are presented that confirm the usefulness of the TSJM for building dynamic prediction models.
The main advantage of the TSJM over MTS is that it avoids bias due to informative dropouts in the first stage. On the other hand, if we disregard parallel computing, the computation time for TSJM is much larger than MTS because a joint model is estimated for each marker at stage 1. Thus, for a limited number of markers, we recommend using existing software for estimating multi-marker joint models such as \texttt{JMBayes2} and \texttt{INLA}. However, when these procedures fail due to an excessive number of markers and random effects, we believe that the proposed TSJM is a viable alternative.
The estimates of joint models obtained by TSJM, as those obtained by MMJM, may be used to compute dynamic individual prediction for any time and any window of prediction.
This is an advantage of the TSJM in comparison with the landmarking approaches for which it is necessary to pre-define the values of the landmark times \cite{ferrer2019individual}.\\
In this paper, we focused on Gaussian markers, but the approach can be extended to non-Gaussian markers. This can be particularly useful for joint models that incorporate both Gaussian and discrete markers, as the estimation of these models often presents convergence issues. 
Also, in the application to 3C data, we do not deal with interval-censoring of dementia. Handling both interval censoring and semi-competing events, using an illness-death model would be a useful extension but considerably increasing the computation time.
We propose
an implementation of the method available at \url{https://github.com/tbaghfalaki/TSJM} combining Bayesian estimation of JM at first stage with JAGS and frequentist estimation of the Cox model with time-dependent variables with the R-package  TSJM. Bayesian approaches provide posterior distributions of the random effects, which are useful for step 2. However, a fully frequentist estimation procedure could also be adopted, partly relying on existing software for each stage, and then implementing the computation of dynamic predictions based on these outputs.\\
{\bf{Acknowledgements}}\\
This work was partly funded by the French National Research Agency (grant ANR-21-CE36 for the project “Joint Models for Epidemiology and Clinical research”). This study was carried out in the framework of the University of Bordeaux's France 2030 program/RRI PHDS.
We thank Christophe Tzourio for providing access to the data from the 3C Study.\\
The 3C Study was supported by Sanofi-Synthélabo, the FRM, the CNAM-TS,DGS,Conseils Régionaux of Aquitaine, Languedoc-Roussillon, and Bourgogne; Foundation of France; Ministry of Research- INSERM ``Cohorts and biological data collections" program; MGEN; Longevity Institute; General Council of the Côte d’Or; ANR PNRA 2006 (grant ANR/ DEDD/ PNRA/ PROJ/ 200206–01-01) and Longvie 2007 (grant LVIE-003-01); Alzheimer Plan Foundation (FCS grant 2009-2012); and Roche.
The Three City Study data are available upon request at e3c.coordinatingcenter@gmail.com.

{\footnotesize{
\bibliographystyle{plain}
\bibliography{biblio.bib}}}

\begin{thebibliography}{10}

\bibitem{abrahamowicz1996time}
Michal Abrahamowicz, Todd MacKenzie, and John~M Esdaile.
\newblock Time-dependent hazard ratio: modeling and hypothesis testing with application in lupus nephritis.
\newblock {\em Journal of the American Statistical Association}, 91(436):1432--1439, 1996.

\bibitem{albert2010estimating}
Paul~S Albert and Joanna~H Shih.
\newblock On estimating the relationship between longitudinal measurements and time-to-event data using a simple two-stage procedure.
\newblock {\em Biometrics}, 66(3):983--987, 2010.

\bibitem{alsefri2020bayesian}
Maha Alsefri, Maria Sudell, Marta Garc{\'\i}a-Fi{\~n}ana, and Ruwanthi Kolamunnage-Dona.
\newblock Bayesian joint modelling of longitudinal and time to event data: a methodological review.
\newblock {\em BMC medical research methodology}, 20:1--17, 2020.

\bibitem{baghfalaki2021approximate}
T~Baghfalaki and M~Ganjali.
\newblock Approximate bayesian inference for joint linear and partially linear modeling of longitudinal zero-inflated count and time to event data.
\newblock {\em Statistical Methods in Medical Research}, 30(6):1484--1501, 2021.

\bibitem{baghfalaki2014joint}
Taban Baghfalaki, Mojtaba Ganjali, and Damon Berridge.
\newblock Joint modeling of multivariate longitudinal mixed measurements and time to event data using a bayesian approach.
\newblock {\em Journal of Applied Statistics}, 41(9):1934--1955, 2014.

\bibitem{beyersmann2011competing}
Jan Beyersmann, Arthur Allignol, and Martin Schumacher.
\newblock {\em Competing risks and multistate models with R}.
\newblock Springer Science \& Business Media, 2011.

\bibitem{blanche2019package}
Paul Blanche.
\newblock Package ‘timeroc’.
\newblock Technical report, updated 2019--12--18. https://cran. r-proj e ct. org/web/packa ges/timeR OC~…, 2019.

\bibitem{blanche2015quantifying}
Paul Blanche, C{\'e}cile Proust-Lima, Lucie Loub{\`e}re, Claudine Berr, Jean-Fran{\c{c}}ois Dartigues, and H{\'e}l{\`e}ne Jacqmin-Gadda.
\newblock Quantifying and comparing dynamic predictive accuracy of joint models for longitudinal marker and time-to-event in presence of censoring and competing risks.
\newblock {\em Biometrics}, 71(1):102--113, 2015.

\bibitem{chi2006joint}
Yueh-Yun Chi and Joseph~G Ibrahim.
\newblock Joint models for multivariate longitudinal and multivariate survival data.
\newblock {\em Biometrics}, 62(2):432--445, 2006.

\bibitem{dafni1998evaluating}
Urania~G Dafni and Anastasios~A Tsiatis.
\newblock Evaluating surrogate markers of clinical outcome when measured with error.
\newblock {\em Biometrics}, 54(4):1445--1462, 1998.

\bibitem{desmee2017nonlinear}
Sol{\`e}ne Desm{\'e}e, France Mentr{\'e}, Christine Veyrat-Follet, Bernard S{\'e}bastien, and J{\'e}r{\'e}mie Guedj.
\newblock Nonlinear joint models for individual dynamic prediction of risk of death using hamiltonian monte carlo: application to metastatic prostate cancer.
\newblock {\em BMC medical research methodology}, 17(1):1--12, 2017.

\bibitem{devaux2022random}
Anthony Devaux, Catherine Helmer, Carole Dufouil, Robin Genuer, and C{\'e}cile Proust-Lima.
\newblock Random survival forests for competing risks with multivariate longitudinal endogenous covariates.
\newblock {\em arXiv preprint arXiv:2208.05801}, 2022.

\bibitem{dutta2021joint}
Srimanti Dutta and Arindom Chakraborty.
\newblock Joint model for longitudinal and time-to-event data: a two-stage approach.
\newblock {\em Journal of Statistics Applications \& Probability}, 10(3):807--819, 2021.

\bibitem{elashoff2016joint}
Robert Elashoff, Ning Li, et~al.
\newblock {\em Joint modeling of longitudinal and time-to-event data}.
\newblock CRC press, 2016.

\bibitem{ferrer2019individual}
Loic Ferrer, Hein Putter, and Cecile Proust-Lima.
\newblock Individual dynamic predictions using landmarking and joint modelling: validation of estimators and robustness assessment.
\newblock {\em Statistical methods in medical research}, 28(12):3649--3666, 2019.

\bibitem{fieuws2004joint}
Steffen Fieuws and Geert Verbeke.
\newblock Joint modelling of multivariate longitudinal profiles: pitfalls of the random-effects approach.
\newblock {\em Statistics in medicine}, 23(20):3093--3104, 2004.

\bibitem{guedj2011joint}
Jeremie Guedj, Rodolphe Thi{\'e}baut, and Daniel Commenges.
\newblock Joint modeling of the clinical progression and of the biomarkers' dynamics using a mechanistic model.
\newblock {\em Biometrics}, 67(1):59--66, 2011.

\bibitem{guler2017two}
Ipek Guler, Christel Faes, Carmen Cadarso-Su{\'a}rez, Laetitia Teixeira, Anabela Rodrigues, and Denisa Mendonca.
\newblock Two-stage model for multivariate longitudinal and survival data with application to nephrology research.
\newblock {\em Biometrical Journal}, 59(6):1204--1220, 2017.

\bibitem{hickey2018joinerml}
Graeme~L Hickey, Pete Philipson, Andrea Jorgensen, Ruwanthi Kolamunnage-Dona, P~Williamson, D~Rizopoulos, and A~Gasparini.
\newblock joinerml: joint modelling of multivariate longitudinal data and time-to-event outcomes.
\newblock {\em R package version 0.4}, 1, 2018.

\bibitem{hossain2020multivariate}
Md~Akhtar Hossain.
\newblock {\em Multivariate Joint Models and Dynamic Predictions}.
\newblock PhD thesis, University of South Carolina, 2020.

\bibitem{huong2018modified}
Pham Thi~Thu Huong, Darfiana Nur, Hoa Pham, and Alan Branford.
\newblock A modified two-stage approach for joint modelling of longitudinal and time-to-event data.
\newblock {\em Journal of Statistical Computation and Simulation}, 88(17):3379--3398, 2018.

\bibitem{ibrahim2001bayesian}
Joseph~G Ibrahim, Ming-Hui Chen, Debajyoti Sinha, JG~Ibrahim, and MH~Chen.
\newblock {\em Bayesian survival analysis}, volume~2.
\newblock Springer, 2001.

\bibitem{kerioui2020bayesian}
Marion Kerioui, Francois Mercier, Julie Bertrand, Coralie Tardivon, Ren{\'e} Bruno, J{\'e}r{\'e}mie Guedj, and Sol{\`e}ne Desm{\'e}e.
\newblock Bayesian inference using hamiltonian monte-carlo algorithm for nonlinear joint modeling in the context of cancer immunotherapy.
\newblock {\em Statistics in Medicine}, 39(30):4853--4868, 2020.

\bibitem{lin2022deep}
Jeffrey Lin and Sheng Luo.
\newblock Deep learning for the dynamic prediction of multivariate longitudinal and survival data.
\newblock {\em Statistics in medicine}, 41(15):2894--2907, 2022.

\bibitem{little1987multiple}
RJA Little and DB~Rubin.
\newblock Multiple imputation for nonresponse in surveys.
\newblock {\em John Wiley \& Sons, Inc.. doi}, 10:9780470316696, 1987.

\bibitem{long2018joint}
Jeffrey~D Long and James~A Mills.
\newblock Joint modeling of multivariate longitudinal data and survival data in several observational studies of huntington’s disease.
\newblock {\em BMC medical research methodology}, 18(1):1--15, 2018.

\bibitem{mauff2020joint}
Katya Mauff, Ewout Steyerberg, Isabella Kardys, Eric Boersma, and Dimitris Rizopoulos.
\newblock Joint models with multiple longitudinal outcomes and a time-to-event outcome: a corrected two-stage approach.
\newblock {\em Statistics and Computing}, 30(4):999--1014, 2020.

\bibitem{mcfetridge2021robust}
Lisa~M McFetridge, {\"O}zg{\"u}r Asar, and Jonas Wallin.
\newblock Robust joint modelling of longitudinal and survival data: Incorporating a time-varying degrees-of-freedom parameter.
\newblock {\em Biometrical Journal}, 63(8):1587--1606, 2021.

\bibitem{mehdizadeh2021two}
P~Mehdizadeh, Taban Baghfalaki, M~Esmailian, and M~Ganjali.
\newblock A two-stage approach for joint modeling of longitudinal measurements and competing risks data.
\newblock {\em Journal of Biopharmaceutical Statistics}, 31(4):448--468, 2021.

\bibitem{moreno2017package}
Margarita Moreno-Betancur, Samuel~L Brilleman, and Maintainer~Margarita Moreno-Betancur.
\newblock Package ‘survtd’.
\newblock {\em https://github.com/moreno-betancur/survtd}, 2017.

\bibitem{moreno2018survival}
Margarita Moreno-Betancur, John~B Carlin, Samuel~L Brilleman, Stephanie~K Tanamas, Anna Peeters, and Rory Wolfe.
\newblock Survival analysis with time-dependent covariates subject to missing data or measurement error: Multiple imputation for joint modeling (mijm).
\newblock {\em Biostatistics}, 19(4):479--496, 2018.

\bibitem{niekerk2021competing}
Janet~van Niekerk, Haakon Bakka, and H{\aa}vard Rue.
\newblock Competing risks joint models using r-inla.
\newblock {\em Statistical Modelling}, 21(1-2):56--71, 2021.

\bibitem{papageorgiou2019overview}
Grigorios Papageorgiou, Katya Mauff, Anirudh Tomer, and Dimitris Rizopoulos.
\newblock An overview of joint modeling of time-to-event and longitudinal outcomes.
\newblock {\em Annual review of statistics and its application}, 6:223--240, 2019.

\bibitem{pintilie2006competing}
Melania Pintilie.
\newblock {\em Competing risks: a practical perspective}.
\newblock John Wiley \& Sons, 2006.

\bibitem{proust2015estimation}
C{\'e}cile Proust-Lima, Viviane Philipps, and Benoit Liquet.
\newblock Estimation of extended mixed models using latent classes and latent processes: the r package lcmm.
\newblock {\em Journal of Statistical Software}, 78(2):1--56, 2017.

\bibitem{proust2009development}
C{\'e}cile Proust-Lima and Jeremy~MG Taylor.
\newblock Development and validation of a dynamic prognostic tool for prostate cancer recurrence using repeated measures of posttreatment psa: a joint modeling approach.
\newblock {\em Biostatistics}, 10(3):535--549, 2009.

\bibitem{rizopoulos2022jmbayes2}
D~Rizopoulos, G~Papageorgiou, and P~Miranda~Afonso.
\newblock Jmbayes2: extended joint models for longitudinal and time-to-event data.
\newblock {\em R package version 0.2-4}, 2022.

\bibitem{rizopoulos2011dynamic}
Dimitris Rizopoulos.
\newblock Dynamic predictions and prospective accuracy in joint models for longitudinal and time-to-event data.
\newblock {\em Biometrics}, 67(3):819--829, 2011.

\bibitem{rizopoulos2012joint}
Dimitris Rizopoulos.
\newblock {\em Joint models for longitudinal and time-to-event data: With applications in R}.
\newblock CRC press, 2012.

\bibitem{rizopoulos2014r}
Dimitris Rizopoulos.
\newblock The r package jmbayes for fitting joint models for longitudinal and time-to-event data using mcmc.
\newblock {\em Journal of Statistical Software}, 72:1--46, 2016.

\bibitem{rizopoulos2011bayesian}
Dimitris Rizopoulos and Pulak Ghosh.
\newblock A bayesian semiparametric multivariate joint model for multiple longitudinal outcomes and a time-to-event.
\newblock {\em Statistics in medicine}, 30(12):1366--1380, 2011.

\bibitem{rizopoulos2014combining}
Dimitris Rizopoulos, Laura~A Hatfield, Bradley~P Carlin, and Johanna~JM Takkenberg.
\newblock Combining dynamic predictions from joint models for longitudinal and time-to-event data using bayesian model averaging.
\newblock {\em Journal of the American Statistical Association}, 109(508):1385--1397, 2014.

\bibitem{rizopoulos2016personalized}
Dimitris Rizopoulos, Jeremy~MG Taylor, Joost Van~Rosmalen, Ewout~W Steyerberg, and Johanna~JM Takkenberg.
\newblock Personalized screening intervals for biomarkers using joint models for longitudinal and survival data.
\newblock {\em Biostatistics}, 17(1):149--164, 2016.

\bibitem{rustand2022denisrustand}
Denis Rustand, Elias~T Krainski, and Haavard Rue.
\newblock Denisrustand/inlajoint: Joint modeling multivariate longitudinal and time-to-event outcomes with inla.
\newblock {\em Github}, 2022.

\bibitem{rustand2022fast}
Denis Rustand, Janet van Niekerk, Elias~Teixeira Krainski, H{\aa}vard Rue, and C{\'e}cile Proust-Lima.
\newblock Fast and flexible inference for joint models of multivariate longitudinal and survival data using integrated nested laplace approximations.
\newblock {\em Biostatistics}, pages 1--20, 2023.

\bibitem{sayers2017joint}
A~Sayers, J~Heron, Andrew~DAC Smith, C~Macdonald-Wallis, MS~Gilthorpe, F~Steele, and K~Tilling.
\newblock Joint modelling compared with two stage methods for analysing longitudinal data and prospective outcomes: a simulation study of childhood growth and bp.
\newblock {\em Statistical methods in medical research}, 26(1):437--452, 2017.

\bibitem{self1992modeling}
Steve Self and Yudi Pawitan.
\newblock Modeling a marker of disease progression and onset of disease.
\newblock {\em AIDS epidemiology: methodological issues}, 1(1):231--255, 1992.

\bibitem{shen2021backward}
Fan Shen and Liang Li.
\newblock Backward joint model and dynamic prediction of survival with multivariate longitudinal data.
\newblock {\em Statistics in medicine}, 40(20):4395--4409, 2021.

\bibitem{song2002estimator}
Xiao Song, Marie Davidian, and Anastasios~A Tsiatis.
\newblock An estimator for the proportional hazards model with multiple longitudinal covariates measured with error.
\newblock {\em Biostatistics}, 3(4):511--528, 2002.

\bibitem{sousa2011review}
In{\^e}s Sousa.
\newblock A review on joint modelling of longitudinal measurements and time-to-even.
\newblock {\em REVSTAT-Statistical Journal}, 9(1):57--81, 2011.

\bibitem{steyerberg2009applications}
Ewout~W Steyerberg.
\newblock Applications of prediction models.
\newblock In {\em Clinical prediction models}, pages 11--31. Springer, 2009.

\bibitem{sweeting2011joint}
Michael~J Sweeting and Simon~G Thompson.
\newblock Joint modelling of longitudinal and time-to-event data with application to predicting abdominal aortic aneurysm growth and rupture.
\newblock {\em Biometrical Journal}, 53(5):750--763, 2011.

\bibitem{sylvestre2008comparison}
Marie-Pierre Sylvestre and Michal Abrahamowicz.
\newblock Comparison of algorithms to generate event times conditional on time-dependent covariates.
\newblock {\em Statistics in medicine}, 27(14):2618--2634, 2008.

\bibitem{taylor2013real}
Jeremy~MG Taylor, Yongseok Park, Donna~P Ankerst, Cecile Proust-Lima, Scott Williams, Larry Kestin, Kyoungwha Bae, Tom Pickles, and Howard Sandler.
\newblock Real-time individual predictions of prostate cancer recurrence using joint models.
\newblock {\em Biometrics}, 69(1):206--213, 2013.

\bibitem{lin2002modeling}
Therneau Terry and Grambsch Patricia.
\newblock Modeling survival data: extending the cox model, 2013.

\bibitem{antoniak2003vascular}
... The 3C Study~Group.
\newblock Vascular factors and risk of dementia: design of the three-city study and baseline characteristics of the study population.
\newblock {\em Neuroepidemiology}, 22(6):316--325, 2003.

\bibitem{therneau2017using}
Terry Therneau, Cindy Crowson, and Elizabeth Atkinson.
\newblock Using time dependent covariates and time dependent coefficients in the cox model.
\newblock {\em Survival Vignettes}, 2(3):1--25, 2017.

\bibitem{therneau2013r}
Terry Therneau and T~Lumley.
\newblock R survival package.
\newblock {\em R Core Team}, 523, 2013.

\bibitem{tsiatis2001semiparametric}
Anastasios~A Tsiatis and Marie Davidian.
\newblock A semiparametric estimator for the proportional hazards model with longitudinal covariates measured with error.
\newblock {\em Biometrika}, 88(2):447--458, 2001.

\bibitem{tsiatis1995modeling}
Anastasios~A Tsiatis, Victor Degruttola, and Michael~S Wulfsohn.
\newblock Modeling the relationship of survival to longitudinal data measured with error. applications to survival and cd4 counts in patients with aids.
\newblock {\em Journal of the American statistical association}, 90(429):27--37, 1995.

\bibitem{van2019joint}
Janet Van~Niekerk, Haakon Bakka, and Haavard Rue.
\newblock Joint models as latent gaussian models-not reinventing the wheel.
\newblock {\em arXiv preprint arXiv:1901.09365}, 2019.

\bibitem{xu2001evaluation}
Jane Xu and Scott~L Zeger.
\newblock The evaluation of multiple surrogate endpoints.
\newblock {\em Biometrics}, 57(1):81--87, 2001.

\bibitem{ye2017two}
Qian Ye and Lang Wu.
\newblock Two-step and likelihood methods for joint models of longitudinal and survival data.
\newblock {\em Communications in Statistics-Simulation and Computation}, 46(8):6019--6033, 2017.

\bibitem{ye2008semiparametric}
Wen Ye, Xihong Lin, and Jeremy~MG Taylor.
\newblock Semiparametric modeling of longitudinal measurements and time-to-event data--a two-stage regression calibration approach.
\newblock {\em Biometrics}, 64(4):1238--1246, 2008.

\bibitem{zhang2023multivariate}
TianHong Zhang, XiaoChen Tang, Yue Zhang, LiHua Xu, YanYan Wei, YeGang Hu, HuiRu Cui, YingYing Tang, HaiChun Liu, Tao Chen, et~al.
\newblock Multivariate joint models for the dynamic prediction of psychosis in individuals with clinical high risk.
\newblock {\em Asian Journal of Psychiatry}, 81:103468, 2023.

\bibitem{zhudenkov2022workflow}
Kirill Zhudenkov, Sergey Gavrilov, Alina Sofronova, Oleg Stepanov, Nataliya Kudryashova, Gabriel Helmlinger, and Kirill Peskov.
\newblock A workflow for the joint modeling of longitudinal and event data in the development of therapeutics: Tools, statistical methods, and diagnostics.
\newblock {\em CPT: Pharmacometrics \& Systems Pharmacology}, 11(4):425--437, 2022.

\end{thebibliography}

\newpage
\section*{Appendix A: Parameter estimation for MMJM  }
\setcounter{equation}{0}
\renewcommand{\theequation}{A.\arabic{equation}}
\setcounter{table}{0}
\renewcommand{\thetable}{A. \arabic{table}}
\setcounter{figure}{0}
\renewcommand{\thefigure}{A. \arabic{figure}}

Let $\bm{\Psi}_k$ and $\bm{\Upsilon}$, respectively, represent the unknown parameters for the $k$th, $k=1,2,\ldots,K$, longitudinal marker model and the time-to-event model, excluding the covariance matrix $\bm{\Sigma}$, and let $\bm{\Psi}=\{\bm{\Psi}_1,\ldots,\bm{\Psi}_K\}$.
If $\bm{y}_{ik}=({y}_{ik1},\ldots,{y}_{ikn_i})^\top$, $\bm{y}_i^\top=(\bm{y}_{i1}^\top,\ldots,\bm{y}_{iK}^\top)^\top$, $\bm{y}^\top=\{\bm{y}_1^\top,\ldots,\bm{y}_N^\top\}$,  $\bm{T}=(T_1,\ldots,T_N)^\top$, $\bm{\delta}=(\delta_1,\ldots,\delta_N)^\top$, the joint likelihood function for models (1) and (3) is as follows:
\begin{eqnarray}\label{like}
 \mathcal{L}(\boldsymbol{\Psi} , \boldsymbol{\Upsilon}, \boldsymbol{\Sigma} \mid \boldsymbol{y}, \boldsymbol{t}, \boldsymbol{\delta})  
& =& \prod_{i=1}^N \int\left(\prod_{k=1}^K \prod_{j=1}^{n_i} \phi\left(y_{i j k} ; \eta_{i k}\left(s_{i j k} \mid \boldsymbol{\beta}_k, \boldsymbol{b}_{i k}\right), \sigma_k^2\right)\right) \\\nonumber
& \times&  \quad \lambda_i\left(t_i \mid \boldsymbol{\omega}_i, \boldsymbol{b}_i, \mathbf{\Upsilon}\right)^{\delta_i} \exp \left(-\Lambda_i\left(t_i \mid \boldsymbol{\omega}_i, \boldsymbol{b}_i, \mathbf{\Upsilon}\right)\right)  \times \phi_p\left(\boldsymbol{b}_i ; \mathbf{0}, \boldsymbol{\Sigma}\right) d \boldsymbol{b}_i,\nonumber
\end{eqnarray}
where  
$\Lambda_{i}(t_i|\bm{\omega}_i,\bm{b}_i,\bm{\Upsilon})=\int_0^{t_i} \lambda_{i}(u|\bm{\omega}_i,\bm{b}_i,\bm{\Upsilon})  du$ and $p=\sum_{k=1}^K p_k$.
Also, $\phi(.;\mu,\sigma^2)$ denotes a univariate normal distribution with mean $\mu$ and variance $\sigma^2$, and
$\phi_p(\bm{b}_i;\bm{0},\bm{\Sigma})$ is the density of a $p$-variate normal distribution with a mean $\bm{0}$ and covariance matrix $\bm{\Sigma}$.\\
In the frequentist paradigm, parameter estimation can be computed by maximizing \eqref{like}, which requires numerical integration of size $p$ \cite{chi2006joint,song2002estimator}. This integration will be more challenging when the dimension of the random effects increases.\\
Another strategy which can be considered for parameter estimation is Bayesian paradigm \cite{xu2001evaluation,rizopoulos2011bayesian,baghfalaki2014joint,long2018joint}. 
In the Bayesian paradigm,  the following joint posterior distribution can be used for parameter estimation: 
\begin{eqnarray}\label{post}
\pi(\boldsymbol{\Psi}, \boldsymbol{\Upsilon}, \boldsymbol{b}, \boldsymbol{\Sigma} \mid \boldsymbol{y}, \boldsymbol{t}, \boldsymbol{\delta}) & \propto & \prod_{i=1}^N\left(\prod_{k=1}^K \prod_{j=1}^{n_i} p\left(y_{i j k} \mid \boldsymbol{b}_{i k}, \boldsymbol{\Psi}_k\right) \times p\left(\boldsymbol{\Psi}_k\right)\right) \\\nonumber
& \times&  \lambda_i\left(t_i \mid \boldsymbol{\omega}_i, \boldsymbol{b}_i, \boldsymbol{\Upsilon}\right)^{\delta_i} \exp \left(-\Lambda_i\left(t_i \mid \boldsymbol{\omega}_i, \boldsymbol{b}_i, \boldsymbol{\Upsilon}\right)\right) \\\nonumber
& \times & \phi_p\left(\boldsymbol{b}_i ; \mathbf{0}, \boldsymbol{\Sigma}\right) \times p(\boldsymbol{\Upsilon}) \times p(\boldsymbol{\Sigma}),\nonumber
\end{eqnarray} 
where $p(\bm{\Psi}_k)$, $ p(\bm{\Upsilon})$ and $p(\bm{\Sigma})$  are the corresponding prior distributions for the unknown parameters. In the simulation studies and applications of the paper, the regression coefficients and association parameters are assumed to follow a zero-mean normal distribution with a large variance as prior distributions. The parameters ${\lambda}_1,\cdots,{\lambda}_J$ are assumed to follow a gamma distribution with a mean of 1 and a large variance. The variance of errors are assumed to have a prior distribution that follows an inverse gamma distribution, while the variance of the random effects is assumed to have a prior distribution that follows an inverse Wishart distribution with degrees of freedom equal to the dimension of the random effects and an identity scale matrix.\\
The exact computation of the joint posterior distribution (\ref{post}) and the other quantities of interest is intractable. There are different practical approaches, such as MCMC \cite{alsefri2020bayesian}, Hamiltonian MC \cite{desmee2017nonlinear,kerioui2020bayesian,mcfetridge2021robust}, or approximate Bayesian inference \cite{van2019joint,niekerk2021competing,baghfalaki2021approximate,rustand2022fast} to make inferences from this kind of posterior distribution. \\
Although using the Bayesian paradigm overcomes the problems of integration in equation \eqref{like} in mathematically straightforward ways, it still faces computational difficulties when the  number of longitudinal markers increases.

\newpage
\section*{Appendix B: Application 1: PBC2 data }
\setcounter{equation}{0}
\renewcommand{\theequation}{B.\arabic{equation}}
\setcounter{table}{0}
\renewcommand{\thetable}{B. \arabic{table}}
\setcounter{figure}{0}
\renewcommand{\thefigure}{B. \arabic{figure}}

\FloatBarrier
\begin{table}[hbt!]
 \caption{\label{pbcest} The estimated regression coefficients of the longitudinal and the time-to-death sub-models from the MMJM and TSJM on PBC2 data. Est: posterior mean, SD: standard deviation, 2.5$\% $ CI: lower bound of credible interval and 97.5$\% $ CI: upper bound of credible interval. }
\centering
\tiny
\begin{tabular}{ccccccccc}
  \hline
 &\multicolumn{4}{c}{TSJM} &\multicolumn{4}{c}{MMJM}  \\    \hline
 & Est. & SD & $2.5\%$ CI & $97.5\%$ CI & Est. & SD & $2.5\%$ CI & $97.5\%$ CI  \\ 
  \hline
  &\multicolumn{8}{c}{Longitudinal sub-model} \\  \hline
   \multicolumn{2}{c}{\bf{Albumin}} &\multicolumn{4}{c}{}  \\    
  Intercept ($\beta_{10}$)  & 3.538 & 0.027 & 3.486 & 3.593 & 3.561 & 0.034 & 3.491 & 3.622 \\ 
  Time ($\beta_{11}$)  & -0.102 & 0.012 & -0.124 & -0.079 & -0.106 & 0.013 & -0.134 & -0.080 \\ 
    $\sigma_1^2$ & 0.093 & 0.004 & 0.085 & 0.102 & 0.094 & 0.004 & 0.086 & 0.102 \\
   \multicolumn{2}{c}{\bf{log(Alkaline)}} &\multicolumn{4}{c}{}  \\    
 Intercept ($\beta_{20}$) & 7.170 & 0.048 & 7.073 & 7.273 & 7.132 & 0.047 & 7.046 & 7.232 \\ 
 Time ($\beta_{21}$) & -0.051 & 0.011 & -0.072 & -0.030 & -0.047 & 0.012 & -0.073 & -0.026 \\ 
   $\sigma_2^2$ & 0.106 & 0.005 & 0.097 & 0.115 & 0.106 & 0.005 & 0.097 & 0.116 \\ 
   \multicolumn{2}{c}{\bf{log(SGOT)}} &\multicolumn{4}{c}{}  \\    
 Intercept ($\beta_{30}$)  & 4.705 & 0.032 & 4.640 & 4.767 & 4.677 & 0.037 & 4.611 & 4.753 \\ 
 Time ($\beta_{31}$)   & -0.002 & 0.011 & -0.024 & 0.019 & 0.007 & 0.011 & -0.013 & 0.029 \\ 
   $\sigma_3^2$ & 0.077 & 0.003 & 0.070 & 0.084 & 0.077 & 0.004 & 0.071 & 0.084 \\ 
   \multicolumn{2}{c}{\bf{log(Platelets)}} &\multicolumn{4}{c}{}  \\    
  Intercept ($\beta_{40}$)  & 5.459 & 0.029 & 5.405 & 5.518 & 5.445 & 0.030 & 5.385 & 5.497 \\ 
  Time ($\beta_{41}$)   & -0.070 & 0.010 & -0.091 & -0.052 & -0.074 & 0.011 & -0.096 & -0.054 \\ 
    $\sigma_4^2$ & 0.043 & 0.002 & 0.039 & 0.047 & 0.043 & 0.002 & 0.039 & 0.047 \\ 
   \multicolumn{2}{c}{\bf{log(SerBilir)}} &\multicolumn{4}{c}{}  \\    
 Intercept ($\beta_{50}$)  & 0.483 & 0.058 & 0.372 & 0.589 & 0.444 & 0.087 & 0.285 & 0.598 \\  
Time ($\beta_{51}$)  & 0.202 & 0.019 & 0.165 & 0.242 & 0.162 & 0.015 & 0.130 & 0.190 \\ 
  $\sigma_5^2$  & 0.115 & 0.006 & 0.105 & 0.126 & 0.112 & 0.005 & 0.102 & 0.123 \\  \hline
    &\multicolumn{8}{c}{Survival sub-model} \\  \hline
 Drug ($\gamma_1$) & -0.044 & 0.199 & -0.421 & 0.359 & -0.132 & 0.233 & -0.589 & 0.327 \\ 
  Age ($\gamma_2$)  & 0.662 & 0.115 & 0.436 & 0.887 & 0.743 & 0.137 & 0.480 & 1.013 \\ 
 Albumin ($\alpha_1$) & -1.613 & 0.278 & -2.182 & -1.088 & -1.779 & 0.358 & -2.489 & -1.115 \\ 
log(Alkaline) ($\alpha_2$) & 0.156 & 0.193 & -0.206 & 0.589 & 0.213 & 0.285 & -0.412 & 0.771 \\ 
 log(SGOT) ($\alpha_3$) & 0.322 & 0.235 & -0.148 & 0.799 & 0.588 & 0.358 & -0.065 & 1.340 \\ 
  log(Platelets) ($\alpha_4$) & -0.371 & 0.210 & -0.804 & 0.035 & -0.604 & 0.286 & -1.153 & -0.094 \\ 
  log(SerBilir) ($\alpha_5$) & 1.216 & 0.128 & 0.957 & 1.462 & 1.038 & 0.172 & 0.691 & 1.365 \\ 
    \hline   
\end{tabular}
\vspace*{10pt}
\end{table}

\newpage
\section*{Appendix C: Tables and figures of application 2}
\setcounter{equation}{0}
\renewcommand{\theequation}{C.\arabic{equation}}
\setcounter{table}{0}
\renewcommand{\thetable}{C. \arabic{table}}
\setcounter{figure}{0}
\renewcommand{\thefigure}{C. \arabic{figure}}

\FloatBarrier
\begin{table}[hbt!]
 \caption{\label{c1} The estimated regression coefficients and variance of errors of the longitudinal sub-model from the joint model of dementia in 3C study data. Est: posterior mean, SD: standard deviation, 2.5$\% $ CI: lower bound of credible interval  and  97.5$\% $ CI: upper  bound of credible interval. }
\centering
\tiny
\begin{subtable}[h]{0.45\textwidth}
        \centering
\begin{tabular}{c|c|cccc}
  \hline
& & Est. & SD & $2.5\%$ CI & $97.5\%$ CI  \\ 
  \hline
SBP & Intercept & 0.169 & 0.117 & -0.037 & 0.384 \\ 
& $Time$ & 0.054 & 0.132 & -0.113 & 0.244 \\ 
& $Time^2$. & -0.324 & 0.238 & -0.522 & 0.014 \\ 
& APOE & 0.017 & 0.077 & -0.128 & 0.171 \\ 
 & Sex & -0.694 & 0.079 & -0.831 & -0.532 \\ 
   & Diabete  & 0.423 & 0.117 & 0.215 & 0.641 \\ 
  & Educ & -0.207 & 0.069 & -0.338 & -0.073 \\ 
   & Age & 0.446 & 0.099 & 0.219 & 0.601 \\ 
& $\sigma^2$ & 1.084 & 0.092 & 0.930 & 1.193 \\  \hline
DBP  & Intercept & 0.462 & 0.109 & 0.231 & 0.664 \\ 
& $Time$ & -0.051 & 0.204 & -0.207 & 0.355 \\ 
& $Time^2$ & -0.051 & 0.098 & -0.289 & 0.009 \\ 
& APOE & -0.061 & 0.077 & -0.206 & 0.091 \\ 
 & Sex & -0.487 & 0.063 & -0.609 & -0.363 \\ 
   & Diabete  & 0.181 & 0.092 & 0.008 & 0.362 \\ 
  & Educ & -0.061 & 0.067 & -0.188 & 0.070 \\ 
   & Age & -0.230 & 0.096 & -0.436 & -0.049 \\ 
& $\sigma^2$ & 0.973 & 0.030 & 0.921 & 1.038 \\  \hline
BMI  & Intercept & 1.378 & 0.291 & 0.852 & 1.932 \\ 
& $Time$ & 0.040 & 0.015 & 0.012 & 0.069 \\ 
& $Time^2$ & -0.003 & 0.002 & -0.006 & 0.002 \\ 
& APOE & -0.472 & 0.285 & -1.174 & 0.023 \\ 
 & Sex & -1.265 & 0.189 & -1.591 & -0.865 \\ 
   & Diabete  & 1.914 & 0.295 & 1.275 & 2.469 \\ 
  & Educ & -0.643 & 0.186 & -0.988 & -0.235 \\ 
   & Age & -0.484 & 0.284 & -0.973 & 0.073 \\ 
& $\sigma^2$ & 0.910 & 0.026 & 0.860 & 0.961 \\  \hline
CESDT  & Intercept & -0.782 & 0.148 & -1.137 & -0.547 \\ 
& $Time$ & 0.121 & 0.232 & -0.055 & 0.614 \\ 
& $Time^2$ & -0.057 & 0.121 & -0.394 & 0.010 \\ 
& APOE & 0.117 & 0.081 & -0.043 & 0.274 \\ 
 & Sex & 0.797 & 0.072 & 0.653 & 0.930 \\ 
   & Diabete  & 0.192 & 0.109 & -0.008 & 0.423 \\ 
  & Educ & -0.163 & 0.063 & -0.296 & -0.045 \\ 
   & Age & 0.418 & 0.125 & 0.163 & 0.677 \\ 
& $\sigma^2$ & 0.999 & 0.035 & 0.943 & 1.083 \\  \hline
BENTON  & Intercept & 0.303 & 0.227 & -0.052 & 0.726 \\ 
& $Time$ & 0.281 & 0.141 & 0.108 & 0.502 \\ 
& $Time^2$ & -0.477 & 0.133 & -0.628 & -0.273 \\ 
& APOE & -0.007 & 0.061 & -0.129 & 0.105 \\ 
 & Sex & -0.171 & 0.062 & -0.283 & -0.046 \\ 
   & Diabete  & -0.172 & 0.088 & -0.348 & -0.015 \\ 
  & Educ & 0.457 & 0.051 & 0.359 & 0.560 \\ 
   & Age & -0.464 & 0.103 & -0.657 & -0.265 \\ 
& $\sigma^2$ & 1.008 & 0.031 & 0.955 & 1.072 \\  \hline
\end{tabular}
\end{subtable}
    \hfill
    \begin{subtable}[h]{0.45\textwidth}
        \centering
\begin{tabular}{c|c|cccc}
  \hline
& & Est. & SD & $2.5\%$ CI & $97.5\%$ CI  \\ 
  \hline
ISA & Intercept & -0.162 & 0.390 & -0.910 & 0.274 \\ 
& $Time$ & 0.220 & 0.202 & 0.057 & 0.649 \\ 
& $Time^2$ & -0.088 & 0.113 & -0.379 & -0.016 \\ 
& APOE & -0.174 & 0.081 & -0.333 & -0.011 \\ 
 & Sex & 0.323 & 0.078 & 0.174 & 0.472 \\ 
   & Diabete  & -0.067 & 0.123 & -0.325 & 0.147 \\ 
  & Educ & 0.818 & 0.082 & 0.669 & 0.996 \\ 
   & Age & -0.703 & 0.184 & -0.991 & -0.358 \\ 
& $\sigma^2$ & 0.943 & 0.031 & 0.885 & 1.012 \\  \hline
  TOTMED & Intercept & -1.173 & 0.228 & -1.646 & -0.841 \\ 
& $Time$ & 0.343 & 0.226 & 0.143 & 0.750 \\ 
& $Time^2$ & -0.085 & 0.134 & -0.425 & 0.000 \\ 
& APOE & 0.115 & 0.100 & -0.069 & 0.322 \\ 
 & Sex & 0.724 & 0.099 & 0.556 & 0.945 \\ 
   & Diabete  & 0.922 & 0.135 & 0.642 & 1.175 \\ 
  & Educ & -0.187 & 0.080 & -0.343 & -0.031 \\ 
   & Age & 0.858 & 0.108 & 0.646 & 1.059 \\ 
& $\sigma^2$ & 0.999 & 0.026 & 0.948 & 1.050 \\  \hline
 TMTA & Intercept & 0.628 & 0.109 & 0.384 & 0.837 \\ 
& $Time$ & 0.113 & 0.148 & 0.034 & 0.562 \\ 
& $Time^2$ & -0.030 & 0.062 & -0.257 & -0.006 \\ 
& APOE & -0.218 & 0.078 & -0.370 & -0.064 \\ 
 & Sex & -0.151 & 0.066 & -0.276 & -0.023 \\ 
   & Diabete  & -0.103 & 0.105 & -0.305 & 0.107 \\ 
  & Educ & 0.342 & 0.068 & 0.210 & 0.472 \\ 
   & Age & -0.760 & 0.098 & -0.949 & -0.545 \\ 
& $\sigma^2$ & 0.939 & 0.040 & 0.880 & 1.051 \\  \hline
 TMTB & Intercept & 0.439 & 0.086 & 0.267 & 0.600 \\ 
& $Time$ & -0.034 & 0.018 & -0.060 & 0.007 \\ 
& $Time^2$ & -0.008 & 0.003 & -0.015 & -0.005 \\ 
& APOE & -0.235 & 0.077 & -0.387 & -0.084 \\ 
 & Sex & -0.166 & 0.075 & -0.307 & -0.016 \\ 
   & Diabete  & -0.235 & 0.119 & -0.492 & -0.012 \\ 
  & Educ & 0.780 & 0.067 & 0.648 & 0.909 \\ 
   & Age & -0.717 & 0.085 & -0.877 & -0.548 \\ 
& $\sigma^2$ & 0.954 & 0.031 & 0.897 & 1.016 \\  \hline
 IADL & Intercept & -0.238 & 0.142 & -0.556 & -0.012 \\ 
& $Time$ & 0.051 & 0.158 & -0.113 & 0.408 \\ 
& $Time^2$ & -0.092 & 0.141 & -0.388 & 0.019 \\ 
& APOE & 0.019 & 0.050 & -0.077 & 0.119 \\ 
 & Sex & 0.084 & 0.048 & -0.009 & 0.176 \\ 
   & Diabete  & 0.065 & 0.063 & -0.057 & 0.188 \\ 
  & Educ & -0.134 & 0.041 & -0.214 & -0.054 \\ 
   & Age & 0.232 & 0.062 & 0.109 & 0.352 \\ 
& $\sigma^2$ & 0.986 & 0.026 & 0.937 & 1.039 \\  \hline
\end{tabular}
\end{subtable}
\vspace*{10pt}
\end{table}
\FloatBarrier

\FloatBarrier
\begin{table}
 \caption{\label{c2} The estimated regression coefficients and variance of errors of the longitudinal sub-model from the joint model of dementia in 3C study data. Est: posterior mean, SD: standard deviation, 2.5$\% $ CI: lower bound of credible interval  and  97.5$\% $ CI: upper  bound of credible interval. }
\centering
\tiny
\begin{subtable}[h]{0.45\textwidth}
        \centering
\begin{tabular}{c|c|cccc}
  \hline
& & Est. & SD & $2.5\%$ CI & $97.5\%$ CI  \\ 
  \hline
WMV  &  Intercept & 2.201 & 0.158 & 1.895 & 2.526 \\ 
& $Time$ & 0.027 & 0.012 & 0.004 & 0.051 \\ 
   & APOE & -0.037 & 0.137 & -0.316 & 0.226 \\ 
  & Sex & -2.601 & 0.113 & -2.826 & -2.380 \\ 
  & Diabete & -0.202 & 0.175 & -0.547 & 0.140 \\ 
   & Educ & 0.265 & 0.111 & 0.043 & 0.475 \\ 
 & Age & -0.919 & 0.148 & -1.198 & -0.617 \\ 
  & $\sigma^2$ & 0.747 & 0.061 & 0.618 & 0.860 \\   \hline
GMW  & Intercept & 6.082 & 0.353 & 5.325 & 6.778 \\ 
& $Time$ & -0.245 & 0.020 & -0.282 & -0.206 \\ 
   & APOE & 0.015 & 0.315 & -0.621 & 0.620 \\ 
  & Sex & -5.023 & 0.257 & -5.522 & -4.534 \\ 
  & Diabete & -1.532 & 0.407 & -2.311 & -0.763 \\ 
   & Educ & 1.327 & 0.249 & 0.831 & 1.789 \\ 
 & Age & -4.736 & 0.320 & -5.342 & -4.131 \\ 
  & $\sigma^2$ & 1.991 & 0.112 & 1.785 & 2.228 \\   \hline
TIV  &  Intercept & 8.582 & 0.684 & 7.297 & 9.873 \\ 
& $Time$ & 0.023 & 0.027 & -0.029 & 0.077 \\ 
   & APOE & -0.439 & 0.662 & -1.658 & 0.915 \\ 
  & Sex & -11.962 & 0.579 & -13.192 & -10.948 \\ 
  & Diabete & -0.955 & 0.776 & -2.399 & 0.611 \\ 
   & Educ & 2.076 & 0.492 & 1.074 & 3.062 \\ 
 & Age & -2.642 & 0.572 & -3.612 & -1.422 \\ 
  & $\sigma^2$ & 3.734 & 0.196 & 3.372 & 4.145 \\   \hline
RHIPP & Intercept & 4.563 & 0.260 & 4.072 & 5.086 \\ 
& $Time$ & -0.316 & 0.013 & -0.341 & -0.290 \\ 
   & APOE & -0.439 & 0.247 & -0.910 & 0.061 \\ 
  & Sex & -2.739 & 0.213 & -3.163 & -2.349 \\ 
  & Diabete & -0.627 & 0.339 & -1.312 & 0.016 \\ 
   & Educ & 0.484 & 0.205 & 0.128 & 0.918 \\ 
 & Age & -3.842 & 0.235 & -4.308 & -3.413 \\ 
  & $\sigma^2$ & 0.899 & 0.062 & 0.779 & 1.026 \\   \hline 
\end{tabular}
\end{subtable}
    \hfill
    \begin{subtable}[h]{0.45\textwidth}
        \centering
\begin{tabular}{c|c|cccc}
  \hline
& & Est. & SD & $2.5\%$ CI & $97.5\%$ CI  \\ 
  \hline
LHIPP & Intercept & 5.269 & 0.282 & 4.682 & 5.811 \\ 
& $Time$ & -0.348 & 0.014 & -0.375 & -0.322 \\ 
   & APOE & -0.288 & 0.273 & -0.843 & 0.215 \\ 
  & Sex & -3.019 & 0.230 & -3.467 & -2.555 \\ 
  & Diabete & -0.935 & 0.360 & -1.604 & -0.185 \\ 
   & Educ & 0.794 & 0.230 & 0.310 & 1.205 \\ 
 & Age & -4.746 & 0.298 & -5.343 & -4.208 \\ 
  & $\sigma^2$ & 1.010 & 0.068 & 0.881 & 1.149 \\   \hline
Peri & Intercept & -1.099 & 0.289 & -1.662 & -0.451 \\ 
& $Time$ & 0.597 & 0.028 & 0.538 & 0.653 \\ 
   & APOE & -0.290 & 0.268 & -0.841 & 0.265 \\ 
  & Sex & -0.615 & 0.207 & -1.002 & -0.199 \\ 
  & Diabete & 0.569 & 0.342 & -0.108 & 1.159 \\ 
   & Educ & 0.264 & 0.235 & -0.178 & 0.736 \\ 
 & Age & 1.773 & 0.285 & 1.184 & 2.315 \\ 
  & $\sigma^2$ & 0.793 & 0.675 & 0.041 & 2.260 \\   \hline
Deep & Intercept & 0.670 & 0.155 & 0.344 & 0.960 \\ 
& $Time$ & 0.019 & 0.012 & -0.007 & 0.043 \\ 
   & APOE & -0.308 & 0.147 & -0.593 & -0.015 \\ 
  & Sex & -0.560 & 0.127 & -0.811 & -0.312 \\ 
  & Diabete & 0.402 & 0.184 & 0.048 & 0.769 \\ 
   & Educ & 0.070 & 0.111 & -0.141 & 0.285 \\ 
 & Age & -0.474 & 0.149 & -0.751 & -0.172 \\ 
  & $\sigma^2$ & 0.558 & 0.123 & 0.247 & 0.755 \\   \hline
\end{tabular}
\end{subtable}
\vspace*{10pt}
\end{table}


\FloatBarrier
\begin{table}
 \caption{\label{c3} The estimated regression coefficients and variance of errors of the longitudinal sub-model from the joint model of death in 3C study data. Est: posterior mean, SD: standard deviation, 2.5$\% $ CI: lower bound of credible interval  and  97.5$\% $ CI: upper  bound of credible interval. }
\centering
\tiny
\begin{subtable}[h]{0.45\textwidth}
        \centering
\begin{tabular}{c|c|cccc}
  \hline
& & Est. & SD & $2.5\%$ CI & $97.5\%$ CI  \\ 
  \hline
SBP & Intercept & 0.094 & 0.169 & -0.213 & 0.372 \\ 
& $Time$ & 0.165 & 0.242 & -0.099 & 0.510 \\ 
& $Time^2$ & -0.112 & 0.148 & -0.431 & 0.011 \\ 
   & APOE & 0.038 & 0.070 & -0.099 & 0.175 \\ 
  & Sex & -0.667 & 0.071 & -0.803 & -0.528 \\ 
  & Diabete & 0.385 & 0.113 & 0.162 & 0.596 \\ 
 & Educ & -0.166 & 0.065 & -0.295 & -0.048 \\ 
 & Age & 0.422 & 0.115 & 0.230 & 0.642 \\ 
  & $\sigma^2$ & 0.947 & 0.022 & 0.905 & 0.990 \\ \hline
DBP  & Intercept & 0.434 & 0.134 & 0.134 & 0.654 \\ 
& $Time$ & 0.055 & 0.173 & -0.217 & 0.297 \\ 
& $Time^2$ & -0.205 & 0.174 & -0.513 & 0.011 \\ 
   & APOE  & -0.032 & 0.076 & -0.184 & 0.121 \\ 
  & Sex & -0.510 & 0.073 & -0.647 & -0.367 \\ 
  & Diabete & 0.191 & 0.103 & -0.011 & 0.395 \\ 
  & Educ & -0.072 & 0.066 & -0.216 & 0.050 \\ 
 & Age & -0.177 & 0.096 & -0.350 & 0.023 \\ 
  & $\sigma^2$ & 0.936 & 0.023 & 0.893 & 0.982 \\ \hline
BMI  & Intercept & 1.221 & 0.240 & 0.804 & 1.707 \\ 
& $Time$ & 0.036 & 0.015 & 0.006 & 0.066 \\ 
& $Time^2$ & -0.002 & 0.002 & -0.006 & 0.001 \\ 
  & APOE & -0.536 & 0.239 & -0.979 & -0.099 \\ 
  & Sex & -1.188 & 0.168 & -1.549 & -0.832 \\ 
   & Diabete & 1.786 & 0.272 & 1.217 & 2.302 \\ 
  & Educ & -0.675 & 0.249 & -1.242 & -0.321 \\ 
 & Age & -0.381 & 0.279 & -0.974 & 0.109 \\ 
  & $\sigma^2$ & 0.916 & 0.026 & 0.868 & 0.970 \\ \hline
CESDT  & Intercept & -0.798 & 0.153 & -1.215 & -0.582 \\ 
& $Time$ & 0.050 & 0.139 & -0.045 & 0.393 \\ 
& $Time^2$ & -0.022 & 0.065 & -0.241 & 0.008 \\ 
  & APOE & 0.114 & 0.079 & -0.044 & 0.263 \\ 
  & Sex & 0.782 & 0.070 & 0.646 & 0.918 \\ 
   & Diabete & 0.222 & 0.096 & 0.029 & 0.402 \\ 
 & Educ & -0.182 & 0.069 & -0.311 & -0.049 \\ 
   & Age & 0.431 & 0.088 & 0.269 & 0.616 \\ 
  & $\sigma^2$ & 0.978 & 0.024 & 0.932 & 1.026 \\ \hline
  BENTON & Intercept & 0.268 & 0.160 & -0.058 & 0.497 \\ 
& $Time$ & 0.156 & 0.206 & -0.038 & 0.498 \\ 
& $Time^2$ & -0.080 & 0.114 & -0.363 & 0.001 \\ 
  & APOE & -0.023 & 0.059 & -0.137 & 0.093 \\ 
    & Sex & -0.179 & 0.048 & -0.275 & -0.086 \\ 
   & Diabete & -0.156 & 0.075 & -0.300 & -0.006 \\ 
 & Educ & 0.439 & 0.049 & 0.345 & 0.538 \\ 
 & Age & -0.478 & 0.067 & -0.598 & -0.337 \\ 
  & $\sigma^2$ & 0.931 & 0.024 & 0.886 & 0.978 \\ \hline
\end{tabular}
\end{subtable}
    \hfill
    \begin{subtable}[h]{0.45\textwidth}
        \centering
\begin{tabular}{c|c|cccc}
  \hline
& & Est. & SD & $2.5\%$ CI & $97.5\%$ CI  \\ 
  \hline
ISA & Intercept & -0.389 & 0.487 & -1.149 & 0.289 \\ 
& $Time$ & 0.403 & 0.314 & 0.042 & 0.877 \\ 
& $Time^2$ & -0.190 & 0.195 & -0.563 & -0.013 \\ 
  & APOE & -0.147 & 0.121 & -0.381 & 0.083 \\ 
   & Sex & 0.330 & 0.088 & 0.146 & 0.487 \\ 
   & Diabete & -0.058 & 0.123 & -0.303 & 0.182 \\ 
 & Educ & 0.824 & 0.091 & 0.662 & 1.007 \\ 
 & Age & -0.717 & 0.120 & -0.949 & -0.497 \\ 
  & $\sigma^2$ & 0.916 & 0.025 & 0.866 & 0.964 \\ \hline
TOTMED & Intercept & -1.170 & 0.158 & -1.464 & -0.890 \\ 
& $Time$ & 0.361 & 0.204 & 0.164 & 0.748 \\ 
& $Time^2$ & -0.114 & 0.139 & -0.389 & -0.004 \\ 
  & APOE & 0.167 & 0.097 & -0.011 & 0.374 \\ 
  & Sex & 0.704 & 0.094 & 0.532 & 0.903 \\ 
  & Diabete & 0.967 & 0.128 & 0.741 & 1.250 \\ 
 & Educ & -0.196 & 0.078 & -0.348 & -0.044 \\ 
 & Age & 0.860 & 0.126 & 0.627 & 1.094 \\ 
  & $\sigma^2$ & 0.986 & 0.024 & 0.939 & 1.034 \\ \hline
TMTA & Intercept & 0.657 & 0.117 & 0.367 & 0.857 \\ 
& $Time$ & 0.079 & 0.095 & 0.022 & 0.418 \\ 
& $Time^2$ & -0.015 & 0.029 & -0.128 & -0.003 \\ 
  & APOE & -0.233 & 0.081 & -0.395 & -0.076 \\ 
 & Sex & -0.142 & 0.066 & -0.269 & -0.012 \\ 
   & Diabete & -0.097 & 0.107 & -0.307 & 0.114 \\ 
 & Educ & 0.323 & 0.066 & 0.196 & 0.451 \\ 
 & Age & -0.803 & 0.088 & -0.976 & -0.635 \\ 
  & $\sigma^2$ & 0.920 & 0.026 & 0.870 & 0.972 \\ \hline
TMTB & Intercept & 0.448 & 0.125 & 0.119 & 0.648 \\ 
& $Time$ & -0.027 & 0.069 & -0.075 & 0.235 \\ 
& $Time^2$ & -0.008 & 0.013 & -0.048 & -0.002 \\ 
  & APOE & -0.262 & 0.096 & -0.443 & -0.054 \\ 
   & Sex & -0.158 & 0.076 & -0.306 & -0.017 \\ 
   & Diabete & -0.192 & 0.102 & -0.380 & 0.027 \\ 
 & Educ & 0.778 & 0.070 & 0.636 & 0.912 \\ 
   & Age & -0.730 & 0.093 & -0.894 & -0.531 \\ 
  & $\sigma^2$ & 0.942 & 0.026 & 0.891 & 0.992 \\ \hline
IADL & Intercept & -0.302 & 0.126 & -0.560 & -0.100 \\ 
& $Time$ & 0.180 & 0.182 & -0.047 & 0.470 \\ 
& $Time^2$ & -0.185 & 0.196 & -0.518 & 0.015 \\ 
  & APOE & 0.009 & 0.049 & -0.090 & 0.105 \\ 
 & Sex & 0.058 & 0.044 & -0.028 & 0.146 \\ 
  & Diabete & 0.084 & 0.066 & -0.041 & 0.212 \\ 
 & Educ & -0.138 & 0.041 & -0.219 & -0.060 \\ 
 & Age & 0.264 & 0.055 & 0.154 & 0.369 \\ 
  & $\sigma^2$ & 0.973 & 0.023 & 0.928 & 1.020 \\ \hline
\end{tabular}
\end{subtable}
\end{table}

\FloatBarrier
\begin{table}
 \caption{\label{c4} The estimated regression coefficients and variance of errors of the longitudinal sub-model from the joint model of death in 3C study data. Est: posterior mean, SD: standard deviation, 2.5$\% $ CI: lower bound of credible interval  and  97.5$\% $ CI: upper  bound of credible interval. }
\centering
\tiny
\begin{subtable}[h]{0.45\textwidth}
        \centering
\begin{tabular}{c|c|cccc}
  \hline
& & Est. & SD & $2.5\%$ CI & $97.5\%$ CI  \\ 
  \hline
  WMV  &  Intercept & 2.190 & 0.144 & 1.900 & 2.462 \\ 
& $Time$ & 0.029 & 0.012 & 0.006 & 0.052 \\ 
   & APOE & -0.030 & 0.131 & -0.291 & 0.226 \\ 
  & Sex & -2.601 & 0.118 & -2.824 & -2.367 \\ 
  & Diabete & -0.228 & 0.175 & -0.566 & 0.127 \\ 
   & Educ & 0.269 & 0.113 & 0.056 & 0.494 \\ 
 & Age & -0.906 & 0.143 & -1.183 & -0.637 \\ 
  & $\sigma^2$ & 0.756 & 0.058 & 0.634 & 0.866 \\   \hline
GMW  & Intercept & 6.005 & 0.370 & 5.250 & 6.719 \\ 
& $Time$ & -0.248 & 0.020 & -0.287 & -0.210 \\ 
   & APOE & 0.018 & 0.299 & -0.597 & 0.607 \\ 
  & Sex & -5.010 & 0.277 & -5.587 & -4.488 \\ 
  & Diabete & -1.556 & 0.401 & -2.351 & -0.765 \\ 
   & Educ & 1.342 & 0.250 & 0.856 & 1.853 \\ 
 & Age & -4.659 & 0.344 & -5.286 & -3.986 \\ 
  & $\sigma^2$ & 1.978 & 0.113 & 1.766 & 2.210 \\   \hline
TIV  &  Intercept & 8.741 & 0.637 & 7.542 & 9.837 \\ 
& $Time$ & 0.018 & 0.027 & -0.036 & 0.070 \\ 
   & APOE & -0.452 & 0.625 & -1.687 & 0.846 \\ 
  & Sex & -12.092 & 0.542 & -13.109 & -11.014 \\ 
  & Diabete & -1.112 & 0.791 & -2.651 & 0.570 \\ 
   & Educ & 2.076 & 0.510 & 1.107 & 3.053 \\ 
 & Age & -2.650 & 0.613 & -3.897 & -1.371 \\ 
  & $\sigma^2$ & 3.718 & 0.189 & 3.368 & 4.121 \\   \hline
RHIPP & Intercept & 4.539 & 0.265 & 4.013 & 5.056 \\ 
& $Time$ & -0.315 & 0.013 & -0.341 & -0.290 \\ 
   & APOE & -0.410 & 0.252 & -0.925 & 0.077 \\ 
  & Sex & -2.703 & 0.233 & -3.162 & -2.266 \\ 
  & Diabete & -0.670 & 0.331 & -1.344 & -0.047 \\ 
   & Educ & 0.535 & 0.206 & 0.122 & 0.936 \\ 
 & Age & -3.888 & 0.238 & -4.349 & -3.409 \\ 
  & $\sigma^2$ & 0.910 & 0.062 & 0.795 & 1.040 \\   \hline
\end{tabular}
\end{subtable}
    \hfill
    \begin{subtable}[h]{0.45\textwidth}
        \centering
\begin{tabular}{c|c|cccc}
  \hline
& & Est. & SD & $2.5\%$ CI & $97.5\%$ CI  \\ 
  \hline
LHIPP & Intercept & 5.278 & 0.300 & 4.671 & 5.860 \\ 
& $Time$ & -0.347 & 0.015 & -0.376 & -0.320 \\ 
   & APOE & -0.208 & 0.303 & -0.757 & 0.430 \\ 
  & Sex & -3.040 & 0.226 & -3.520 & -2.616 \\ 
  & Diabete & -0.940 & 0.342 & -1.623 & -0.282 \\ 
   & Educ & 0.754 & 0.228 & 0.301 & 1.194 \\ 
 & Age & -4.759 & 0.292 & -5.327 & -4.190 \\ 
  & $\sigma^2$ & 1.012 & 0.069 & 0.884 & 1.154 \\   \hline
Peri & Intercept & -1.001 & 0.272 & -1.504 & -0.489 \\ 
& $Time$ & 0.594 & 0.030 & 0.542 & 0.654 \\ 
   & APOE & -0.239 & 0.351 & -0.952 & 0.414 \\ 
  & Sex & -0.628 & 0.230 & -1.049 & -0.118 \\ 
  & Diabete & 0.440 & 0.509 & -0.441 & 1.252 \\ 
   & Educ & 0.141 & 0.232 & -0.302 & 0.577 \\ 
 & Age & 1.741 & 0.258 & 1.227 & 2.259 \\ 
  & $\sigma^2$ & 1.183 & 1.125 & 0.018 & 3.627 \\   \hline
Deep & Intercept & 0.658 & 0.165 & 0.330 & 0.984 \\ 
& $Time$ & 0.019 & 0.012 & -0.005 & 0.042 \\ 
   & APOE & -0.278 & 0.143 & -0.560 & 0.005 \\ 
  & Sex & -0.556 & 0.125 & -0.789 & -0.312 \\ 
  & Diabete & 0.441 & 0.182 & 0.085 & 0.797 \\ 
   & Educ & 0.061 & 0.126 & -0.171 & 0.324 \\ 
 & Age & -0.470 & 0.145 & -0.744 & -0.185 \\ 
  & $\sigma^2$ & 0.565 & 0.113 & 0.297 & 0.759 \\ 
   \hline   
\end{tabular}
\end{subtable}
\end{table}
\FloatBarrier

\begin{figure}[h]
\centering
\includegraphics[width=10cm]{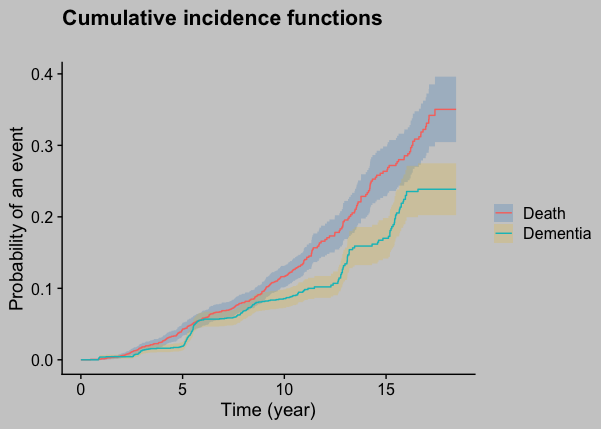}
\vspace*{-.1cm} \caption{\label{cif} Cumulative incidence function for dementia and death event over 17 years of follow-up. }
\end{figure}
\FloatBarrier

\begin{landscape}
\begin{figure}[h]
\centering
\includegraphics[width=20cm]{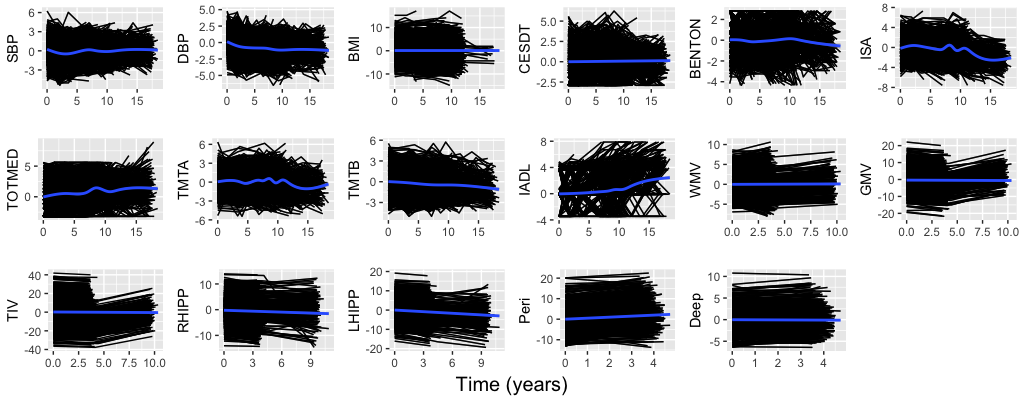}
\vspace*{-.1cm} \caption{\label{sp} Individual trajectories from the normalized clinical, neuropsychological and imaging longitudinal markers in the 3C study. Systolic blood pressure (SBP), 
diastolic  blood pressure (DBP), body mass index (BMI),  depressive symptomatology  measured  
using the Center for Epidemiologic Studies Depression scale  (CESDT), the visual retention test of Benton (BENTON), Isaac Set Test (ISA), the total number of medications (TOTMED), the trail making test A and B (TMTA and TMTB), functional dependency assessed using Instrumental Activity of Daily Living scale (IADL),
white matter volume (WMV), gray matter volume (GMW),  total intracranial volume  (TIV),
 right and left hippocampal volume (RHIPP and LHIPP), 
for the 5-year risk of dementia of thevolumes of White Matter Hyperintensities in the periventricular (Peri) and deep (Deep) white matter.}
\end{figure}
\end{landscape}
\FloatBarrier

\end{document}